\documentclass[journal]{IEEEtran}

\usepackage[ruled,noline]{algorithm2e}
\usepackage{algorithmic}
\usepackage{setspace}
\usepackage{amssymb}
\usepackage{amsmath}
\usepackage{latexsym}
\usepackage{framed}
\usepackage{color, soul}
\usepackage{xargs}
\usepackage{tikz}
\usepackage{todonotes}
\usepackage{multirow} 
\usepackage{verbatim}

\DeclareMathOperator*{\argmin}{\textit{argmin}}
\DeclareMathOperator{\E}{\mathbb{E}}

\usepackage{graphicx}
\graphicspath{{../pdf/}{../jpeg/}}
\DeclareGraphicsExtensions{.pdf,.jpeg,.png}

\newcommand{\bumpup}{\vspace*{-2.0ex}}

\usepackage{array}
\usepackage{mdwtab}
\usepackage{eqparbox}
\usepackage{url}
\begin{document}

\title{PSIGAN: Joint probabilistic segmentation and image distribution matching for unpaired cross-modality adaptation based MRI segmentation}
\author{Jue Jiang, Yu Chi Hu, Neelam Tyagi,  Andreas Rimner, Nancy Lee, Joseph O. Deasy, Sean Berry, Harini Veeraraghavan
	\thanks{This work was supported by the MSK Cancer Center core grant P30 CA008748.}
	
	\thanks{Jue Jiang, Neelam Tyagi, Yu Chi-Hu Hu, Joseph O. Deasy, Sean Berry, and Harini Veeraraghavan are with the department of Medical Physics, Memorial Sloan Kettering Cancer Center, New York, USA(email: veerarah@mskcc.org)}
	\thanks{ Nancy Lee and Andreas Rimner are with the department of Radiation oncology, Memorial Sloan Kettering Cancer Center, New York, USA.}%
}	
\maketitle            % typeset the header of the 

\begin{abstract}
We developed a new joint probabilistic segmentation and image distribution matching generative adversarial network (PSIGAN) for \textcolor{black}{unsupervised} domain adaptation (UDA) and multi-organ segmentation from magnetic resonance (MRI) images. Our UDA approach models the co-dependency between images and their segmentation as a joint probability distribution using a new structure discriminator. The structure discriminator computes structure of interest focused adversarial loss by combining the generated pseudo MRI with probabilistic segmentations produced by a simultaneously trained segmentation sub-network. The segmentation sub-network is trained using the pseudo MRI produced by the generator sub-network. This leads to a cyclical optimization of both the generator and segmentation sub-networks that are jointly trained as part of an end-to-end network.  Extensive experiments and comparisons against multiple state-of-the-art methods were done on four different MRI sequences totalling 257 scans for generating multi-organ and tumor segmentation. The experiments included, (a) 20 T1-weighted (T1w) in-phase mdixon and (b) 20 T2-weighted (T2w) abdominal MRI for segmenting liver, spleen, left and right kidneys, (c) 162 T2-weighted fat suppressed head and neck MRI (T2wFS) for parotid gland segmentation, and (d) 75 T2w MRI for lung tumor segmentation. Our method \textcolor{black}{achieved} an overall average DSC of 0.87 on T1w and 0.90 on T2w for the abdominal organs, 0.82 on T2wFS for the parotid glands, and 0.77 on T2w MRI for lung tumors. We have made our code available at: \textcolor{blue}{https://github.com/harveerar/PSIGAN}.

\end{abstract}

\section{Introduction}
\IEEEPARstart{M}{agnetic} resonance imaging (MRI) is rapidly emerging as the modality for image-guided adaptive radiation therapy treatments\cite{kupelian2014} due to its better soft tissue contrast compared with computed tomography (CT) scans. However, a critical requirement for MR-guided radiotherapy is fast, accurate, and consistent segmentation of target and surrounding normal organs at risk (OAR)~\cite{bainbridge2017}. 
\begin{figure}
\centering
\includegraphics[width=0.4\textwidth]{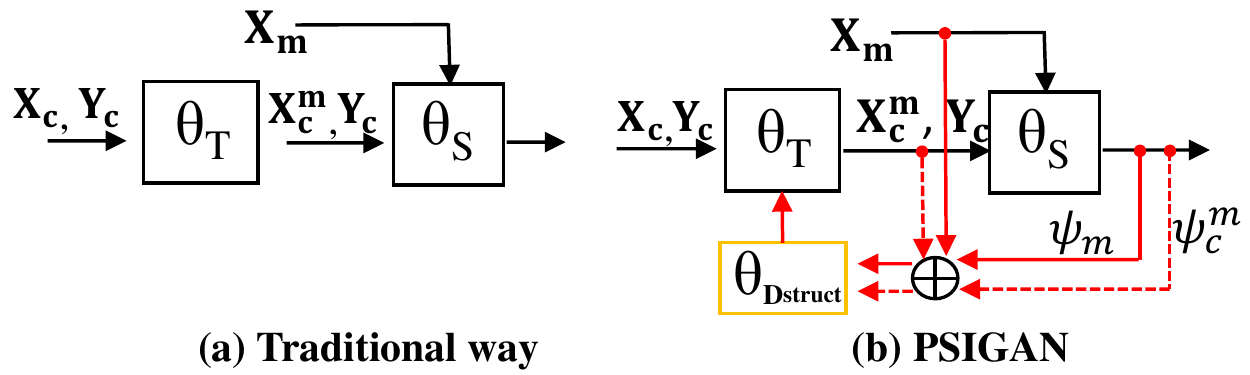}
\caption{\small{Difference between traditional UDA \cite{huo2018synseg} and PSIGAN segmentation. In both (a) and (b), the translation network $T$ parameterized by $\theta_{T}$ produces labeled pseudo target data $\{X_{c}^{m}, Y_{c}\}$ from unpaired source $X_c$ and target $X_m$ data and trains the segmentation (S) network, parameterized by $\theta_{S}$. PSIGAN also uses a structure discriminator $D_{struct}$ to match the joint distribution of image-segmentation probability maps $\{x_m,\psi_m\},\{x_c^m,\psi_c^m\}$ to further optimize $\theta_T$. \label{fig1:method_schematic}}}
\end{figure} 

Deep learning-based methods have shown remarkable success in diverse image analysis tasks when they can be trained using large  annotated datasets. However, acquiring large expert-segmented medical image datasets is difficult. This is because, slice-by-slice delineations of several organs is highly time consuming, and such delineations require a domain expert, such as a radiologist or a radiation oncologist. 

Domain adaptation~\cite{kamnistas2017,zhuTMI2019,Zhang_2018_CVPR,ouyangMICCAI2019} is a commonly used approach to overcome the  issue of learning from limited and unlabeled target modality datasets, where a model for the target domain is trained by using an existing labeled dataset from a different modality, called the source domain. In unsupervised domain adaptation (UDA) based segmentation, the focus of this work, no target domain labeled data is available for training. 

UDA segmentation has been accomplished by either using feature-level or pixel-level adaptation. In feature-level adaptation, the encoder networks are adversarially trained to extract domain-invariant features, such that a single segmentation model trained on source data is applicable to both domains~\cite{kamnistas2017,zhuTMI2019,li2019bidirectional,duo2019IEEEAccess}. In  pixel-level adaptation, generative adversarial networks (GAN) model the complex inter-modality anatomical relationships and compute image to image (I2I) translations~\cite{Zhang_2018_CVPR,jiang2018tumor}. The target modality model is then learned by using the source to target transformed images. Cyclical consistency~\cite{zhu2017unpaired} and feature disentanglement~\cite{lee2018diverse,yang2019unsupervised} losses are often used in unpaired I2I translation methods to circumvent the lack of corresponding source and target modality images~\cite{huo2018synseg,zhang2018,ouyangMICCAI2019}. Hybrid methods combine feature and pixel-level adaptation~\cite{bousmalis2017unsupervised,hoffman2017cycada,chen2019synergistic} in order to ensure good pixel-level I2I transformations and the preservation of low-level edge and mid-level textural characteristics of the target modality images. 

A major issue while performing UDA is mode collapse~\cite{salimans2016}, wherein multiple distinct inputs are mapped to a same output. Modality hallucination is a related manifestation of this issue that commonly occurs in medical image I2I translations, wherein distinct organ characteristics like geometry (or overall shape) and appearance (or intensity distribution) are ill-preserved or are removed~\cite{jiang2018tumor} in translated images. This is because the commonly used losses based on matching global or marginal intensity statistics of whole images cannot sufficiently constrain the generator to model the local organ/tumor geometry and appearance statistics. Feature disentanglement methods have been reported to reduce the \textcolor{black}{aforementioned} issues by using shared content  and domain-specific attribute encoders~\cite{lee2018diverse}, with demonstrated success in medical image applications~\cite{yang2019unsupervised}. Nevertheless, the lack of explicit conditioning of losses with respect to the \textcolor{black}{geometry} and \textcolor{black}{appearance} of the various structures of interest (SOI), may not yield the desired results for the output task.

Prior works have used output segmentation as an adversarial loss to constrain UDA and improve segmentations~\cite{zhang2018,cheng2018,tsai2019domain,vu2019advent,jiang2018tumor}. As shown in~\cite{bousmalis2017unsupervised},  task-specific losses can improve training stability and reduce chances of mode collapse. However, adversarial losses computed only using the segmentation output can constrain the \textcolor{black}{geometry but not necessarily the appearance of SOIs in I2I translation}. \par
A key difference of our approach compared to prior works is the use of joint distribution matching adversarial loss. \textcolor{black}{The joint distribution is represented as a channel-wise concatenation of images and their voxel-wise segmentation probability maps.} Prior works have employed similar joint-distribution matching to constrain bi-directional mapping between images and a low dimensional latent distribution vector~\cite{li2017ALICE,donahue2016adversarial} or images and scalar output categories~\cite{chongxuan2017triple}. \textcolor{black}{Ours on the other hand constrains pixel-level relationships between I2I translations and segmentations.} %The use of segmentation probability maps instead of a binary segmentation allows the discriminator to incorporate the uncertainties (\textcolor{black}{e.g. larger uncertainty in the SOI-background interface vs center of SOI)} in the segmentation when computing domain mismatches.  \\
To our best knowledge, this is the first approach to compute joint distribution matching of images and segmentation probabilities for UDA  segmentation (Fig.~\ref{fig1:method_schematic}). This approach also leads to a cyclical optimization where both segmentation and generator outputs constrain each other. In prior approaches like~\cite{vu2019advent,li2019bidirectional}, segmentation outputs do not constrain generator network gradient computation.

Our contributions are:
\begin{itemize}
	\item \textit{Organ geometry and appearance constrained unpaired cross-modality adaptation.} We introduced an UDA approach that constrains organ geometry and appearance in I2I translations by computing adversarial losses to minimize the mismatch in the joint distribution of images and their segmentation probability maps. 
	\item \textit{A cyclic feedback based UDA segmentation.} In our approach, the generator outputs are used to train the segmentation network, while the segmentation outputs are used to compute losses for the generator.  	%\textcolor{black}{(I fee this optimization is related to wake-sleep algorithm, or maybe can put in discussion.)}\textcolor{red}{Not really.}
	%combines pairs of target image and its segmentation similar to the segmentation probability maps with  corresponding synthesized or target image to measure the joint density matching of the \textcolor{black}{(pair of real image with it's segmentation probability map and pair of pseudo image with it's segmentation map )} target and segmentation probability maps. The segmentation probability maps \textcolor{black}{containing structure information} are used like attention maps in attention-guided translation\cite{mejjati2018unsupervised,zhang2019ICML} but \textcolor{red}{with no additional network parameters required for computing attention and scales our approach to model multiple structures transformation}. 
	\item \textit{Comprehensive performance comparisons \/}\rm were done against multiple state-of-the-art methods using three datasets for multiple organs and tumor segmentation. \textcolor{black}{Extensive ablation experiments were done to measure the impact of joint distribution matching on both segmentation and I2I translation.} 
\end{itemize}
\begin{figure*}
\centering
\includegraphics[width=0.875\textwidth]{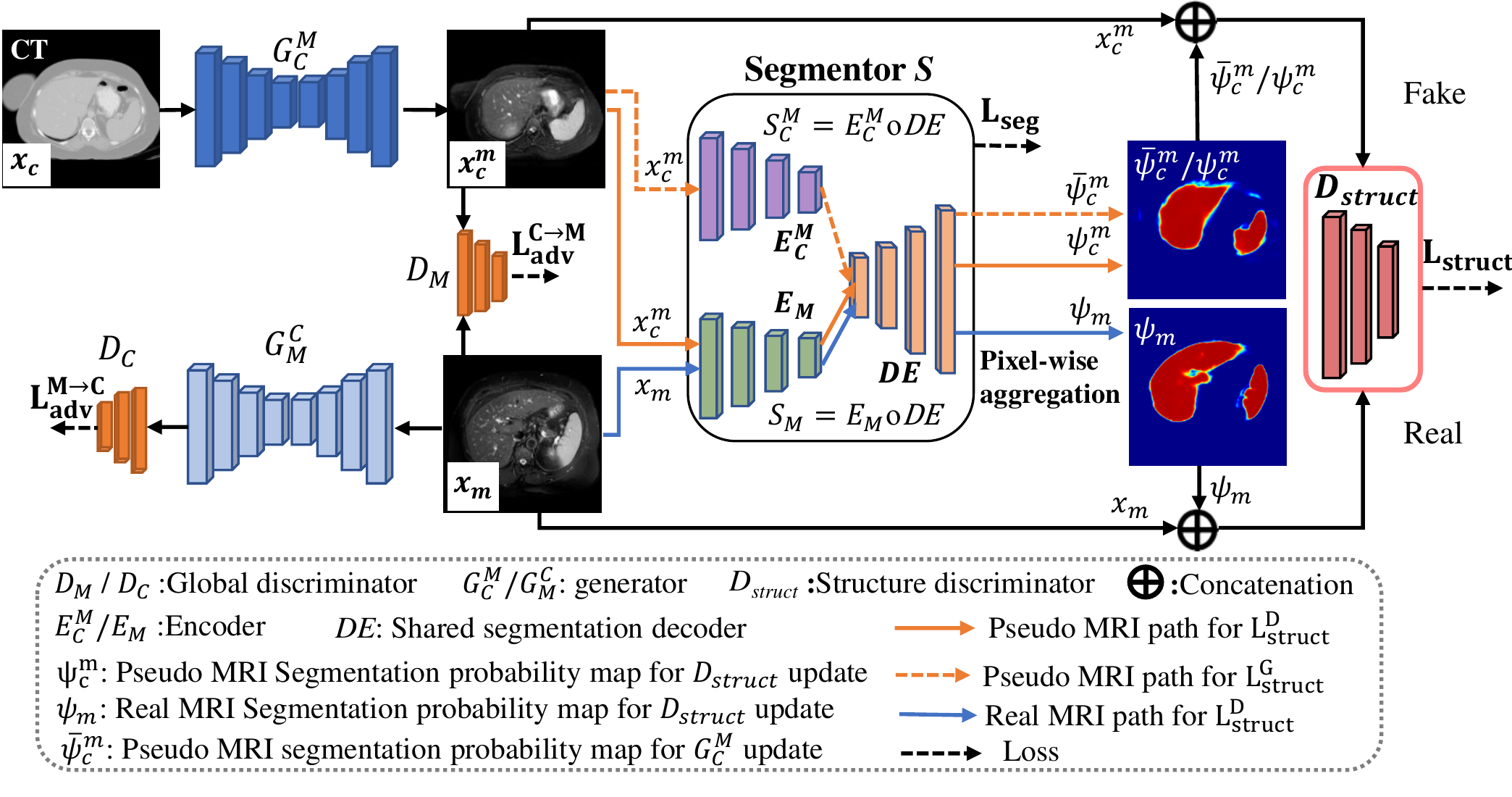}
\caption{\small{\textcolor{black}{Approach overview. Generator $G_C^M$ converts CT image $x_c$ into pseudo MR image $x_c^m$, which is used to train the segmentor $S$. $S$ uses split encoders $E_M, E_C^M$ and shared decoder $DE$ to separate the gradient flows for  $D_{struct}$ and $G_C^M$ through sub-networks $S_M$ and $S_C^M$. $D_{struct}$ computes joint distribution matching ($\{x_c^m, \psi_c^m\}$, $\{x_m, \psi_m\}$) adversarial loss $L_{struct}^D$, where $\psi_c^m$ and $\psi_m$ are produced by $S_M$. The corresponding  adversarial loss $L_{struct}^G$ for $G_C^M$ uses $\{x_c^m, \Bar{\psi}_c^m\}$, where $\Bar{\psi}_c^m$ is produced by $S_C^M$. $D_M$ and $D_C$ are global \textcolor{black}{intensity } discriminators for MRI and CT domain; $G_M^C$ converts MRI to CT images for enforcing cyclically consistent I2I translations.}}
\label{fig1:method_overview}
}

\end{figure*}

\section{Related Works}
\subsection{Feature-level UDA \textcolor{black}{segmentation}}
Feature-level UDA segmentation methods extract a domain-invariant feature encoding, such that a model trained on source data is applicable to both source and target domains. This method is often used for performing domain adaptation between related image sequences. Example applications of this include prostate gland segmentation from MRIs acquired from different scanners~\cite{zhuTMI2019}, and brain tumor segmentation from similar MRI contrasts~\cite{kamnistas2017}. \textcolor{black}{Joint adversarial training strategy combining a domain critic network with a segmentation network has been used for the more challenging cross-modality adaptation between CT and MRI in~\cite{dou2018unsupervised}.} However, as shown in~\cite{dou2018unsupervised}, these methods may require different number of feature layers to be adapted for segmenting various organs. Hence, network feature-adaptation depth is an important hyper-parameter. Such a depth tuning can require computationally intensive training when generating segmentation of large number of organs for radiation therapy applications. This issue has been avoided by using joint latent space learning with variational autoencoders~\cite{ouyangMICCAI2019} and disentanged feature representations~\cite{joyce2018deep}. The work in~\cite{dong2018unsupervised} employed output feature matching for producing scanner invariant estimation of cardiothoracic ratio (or heart size) from chest X-rays. Segmentation probability maps produced from \textit{softMax} layer~\cite{tsai2018learning} and entropy maps indicative of pixel-wise classification uncertainties~\cite{vu2019advent} have been used for domain invariant (synthetic and camera-acquired) natural image segmentation.

\subsection{Pixel-level UDA \textcolor{black}{segmentation}}
\textcolor{black}{Pixel-level domain adaptation and segmentation methods model the inter-modality relationships and compute I2I image translations.} The translated images are used to train a segmentation model for the target domain~\cite{murez2018image,liu2019susan,huo2018synseg,zhang2018}. Two-step training consisting of I2I translation (e.g., CT to MRI) followed by segmentation network training has been used for MRI lung tumor~\cite{jiang2018tumor} and fundus image segmentation~\cite{zhao2018supervised}. I2I translation and segmentation were combined into one network to segment cardiac structures from CT~\cite{zhang2018,chen2019synergistic}, abdominal organs from CT and MRI~\cite{huo2018synseg}, and knee structures from MRI~\cite{liu2019susan}.
\textcolor{black}{Multiple works~\cite{cai2018towards,liu2019susan,huo2018synseg} have combined cyclical consistency losses of the CycleGAN~\cite{zhu2017unpaired} with a segmentation network to train with unpaired source and target modalities. Cyclical consistency loss shrinks the space of possible mappings in GANs by computing global intensity distribution mismatches.} But this loss alone is insufficient to preserve the SOI geometry and appearance in I2I translation of medical images~\cite{cohen2018}. Inclusion of style and perceptual losses have shown improvements in I2I translations~\cite{armanious2018}. Geometry preserving losses implemented by \textcolor{black}{backpropagating segmentation losses~\cite{zhang2018} to the generator} and high-level segmentation features matching \textcolor{black}{losses} have also shown improvements for both tumor~\cite{jiang2018tumor} and semantic segmentation of real-world images~\cite{hoffman2017cycada}. However, none of the \textcolor{black}{aforementioned} losses provide constraints to sufficiently control both geometry and appearance of the SOIs in I2I translation. We address this problem by combining images and their segmentation probability maps as a joint density for adversarial learning.

\section{Method}
\textcolor{black}{\textbf{Goal: \/}Learn MRI multi-organ segmentation models by using unpaired expert-segmented CT and unlabeled MRI images.}\par
\textbf{Notation:\/} CT $(X_{c}, Y_{c})$ is the source domain that consists of images $x_c \in X_C$ and expert-segmentations $y_c \in Y_C$ for training. MRI is the target domain that is provided with only MRI images $x_m \in X_M$ for training. \textcolor{black}{As the learning optimization involves a finite set of examples, the probability distribution of CT and MRI images are represented as $p(x_c)$ and $p(x_m)$. The probability distribution of pseudo MRI $x_c^m$ and pseudo CT $x_m^c$ resulting from I2I translations of CT to MRI and MRI to CT are represented as $p(x_c^m)$ and $p(x_m^c)$, respectively. The joint probability distribution of the real MR and pseudo MR images and their probabilistic segmentations are represented as $p(x_m, \psi_m)$ and $p(x_c^m,\psi_c^m)$, respectively.} 
\bumpup
\subsection{Background}
In supervised learning with finite set of training examples \textcolor{black}{$N$}, given a joint probability distribution of inputs and outputs $p(x,y)$ and a model parameterized by $\theta$, a chosen loss function $L_c(.)$ is used to compute the empirical risk in predicting the outputs $y$ from inputs $x$: 
\begin{equation}
\setlength{\abovedisplayskip}{1pt}
\setlength{\belowdisplayskip}{1pt}
\E[L(x,y,\theta)] = \argmin_{\theta \in \Theta} \sum_{i=1}^{N} p(x_i,y_i) L_c(x_{i},y_{i},\theta),
\end{equation}

In pixel-level UDA segmentation, we seek to minimize the empirical risk of training with pseudo target examples, \textcolor{black}{obtained through a model $\phi:x_{c} \rightarrow x_{c}^{m}$, by assuming that $p(x_c^m) \approx p(x_m)$:} %~\footnote{This representation can be image or feature map. Here, we use image.} 
\begin{equation}
\setlength{\abovedisplayskip}{1pt}
\setlength{\belowdisplayskip}{1pt} 
\textcolor{black}{
 \E[L(x_c^m,y,\theta)] = \argmin_{\theta \in \Theta} \sum_{i=1}^{N} p(x_{c_{i}}^m,y_i) L_c(x_{c_{i}}^m,y_{i},\theta).}
 \label{eqn:EMR_UDA}
\end{equation}
\textcolor{black}{where $p(x_c^m,y)$ is the joint distribution on an intermediate representation or pseudo domain. However, $\phi$ does not produce a perfect  mapping, that is $p(x_c^m) \neq p(x_m)$. Therefore, an additional domain translation loss $L_t$ needs to be added to (\ref{eqn:EMR_UDA}). With $p(x_c^m,x_m)$ as the joint probability distribution over pseudo and real target samples, the risk is computed as:} 
\begin{equation}
\begin{split}
\textcolor{black}{
\E[L(x_{c}^{m},x_{m},y,\theta,\phi)]} = & \textcolor{black}{\underbrace{\argmin_{\phi \in \Phi} \sum_{j=1}^{M} p(x_{c_{j}}^{m},x_{m_{j}}) L_t(x_{c_{j}}^{m},x_{m_{j}},\phi)}_\text{Domain translation}}  \\ & + \textcolor{black}{\underbrace{\argmin_{\theta \in \Theta} \sum_{i=1}^{N} p(x_{c_{i}}^{m},y_i) L_c(x_{c_{i}}^{m},y_{i},\theta)}_\text{Segmentation}. 
}
\end{split}
 \label{eqn:EMR_UDA_expand}
\end{equation}
Equation (\ref{eqn:EMR_UDA_expand}) can be optimized by training the domain translation and  segmentation networks either sequentially~\cite{jiang2018tumor,zhao2018supervised} or jointly~\cite{huo2018synseg,zhang2018,hoffman2017cycada}. However, this optimization ignores the co-dependency of domain translation and segmentation. For instance, no explicit constraint exists to preserve any inherent target modality appearance or geometric characteristics of the SOIs that distinguishes them from the background in the I2I translated images, which is a cause of  sub-optimal performance. We model this co-dependency by computing adversarial losses using the joint distribution of images and their segmentation probability maps as a pair obtained from the translated ($\{x_{c}^{m},\psi_{c}^{m}\}$) and real target images ($\{x_{m},\psi_{m}\}$). In other words, we require $p(x_c^m,\psi_c^m) \approx p(x_m, \psi_m)$.
\subsection{Probabilistic segmentation and image matching GAN (PSIGAN)}

\textcolor{black}{The overview of PSIGAN is shown in Fig.~\ref{fig1:method_overview}}. PSIGAN consists of a CT to MRI generator $G_C^M$:$x_{c} \rightarrow x_c^m$,  global \textcolor{black}{intensity} discriminator $D_{M}$, a structure discriminator $D_{struct}$, and a target domain segmentor $S$: $x_{m}\rightarrow \{\psi_m, y_m\}$, where $\psi_{m}$ is the predicted map of voxel-wise segmentation probabilities for $x_m$ and $y_m$ is the K-organ segmentation for $x_m$.  \textcolor{black}{We also include a MR to CT generator $G_M^C$:$x_m \rightarrow x_m^c$ and global \textcolor{black}{intensity} discriminator $D_C$ to implement cyclical consistency losses for unpaired I2I translation. $D_{M}$ computes a global adversarial loss to penalize mismatches in the marginal intensity distribution of pseudo and real MRI images in order that $p(x_c^m) \approx p(x_m)$:}
\begin{equation}
\label{eqn:over_adversary_loss_M}
\begin{split}
%\scalemath{0.7}
& \textcolor{black}{\max\limits_{D_M}\min\limits_{G_C^M}} \ L^{C\rightarrow M}_{adv}(G_{C}^{M}, D_{M})=\E_{x_m \sim p(x_m)} [log(D_{M}(x_m))] +  \\ & \textcolor{black}{\E_{x_c \sim p(x_c)} [log(1-(D_{M}(G_{C}^{M}(x_c))]}. 
%& L^{C\rightarrow M}_{adv}(G_{C}^{M}, D_{C}^{M}, X_{M}, X_{C})= \E_{x_{m} \sim X_{m}} [log(D_{C}^{M}(x_{m}))] + \\ & \E_{x_{c} \sim X_{C}} [log(1-(D_{C}^{M}(G_{C}^{M}(x_{c}))]. 
\end{split} 
\end{equation}

\textcolor{black}{$D_{struct}$ computes a joint distribution adversarial loss to reduce SOI appearance (intensity distribution) and geometry (overall shape) mismatches between pseudo $x_c^m$ and real MRIs $x_m$. It accomplishes this by matching the joint distributions $p(x_m, \psi_m)$ and $p(x_c^m, \psi_c^m)$, implemented by concatenating the images and their segmentation probability maps. The segmentor $S$ (Fig.~\ref{fig1:method_overview})} produces the segmentation probability maps $\psi_m$, $\psi_c^m$ from real $x_m$ and pseudo MRI $x_c^m$. $D_{struct}$ loss is computed by using both real and pseudo MRI as:
\begin{equation}
\label{eqn:Joint_D_psi}
\begin{split}
&\textcolor{black}{\max\limits_{D_{struct}}}\textcolor{black}{L_{struct}^D= \E_{(x_m,\psi_m) \sim p(x_m,S(x_m))}} [log(D_{struct}(x_m, \psi_m))] \\ & + 
\textcolor{black}{\E_{(x_c^m, \psi_c^m) \sim p(x_c^m,S(x_c^m))}} [log(1-D_{struct}(x_c^m,\psi_c^m))]. 
\end{split} 
\end{equation}   
 
\textcolor{black}{The inclusion of segmentation probability maps in this loss constrains the SOI geometry similar to prior works~\cite{huo2018synseg,zhang2018,zhao2018supervised,chen2019synergistic}. The inclusion of images in this loss additionally constrains the appearance.}  
\textcolor{black}{The generator $G_C^M$ \textcolor{black}{is optimized by} adversarial loss, \textcolor{black}{which} is computed using the pseudo MR images $x_c^m$\textcolor{black}{ = $G_C^M$($x_c$)} as:} 
\begin{equation}
\begin{split}
&\textcolor{black}{\min\limits_{G_C^M}} \ L_{struct}^G = \\ & \textcolor{black}{\E_{(x_c^m, \psi_c^m) \sim p(x_c^m,S(x_c^m))}} [log(1-D_{struct}(x_c^m,\psi_c^m))].
\end{split} 
\label{eqn:Joint_G_psi}
\end{equation}    
$\psi_m, \psi_c^m$ are produced by \textcolor{black}{aggregating label assignment probabilities $e_{i,j}$ at location $i,j$ for $K$ SOIs from $K$ channels as generated by the \textit{softMax \/}\rm layer of the network S as:} 
\begin{equation}
\begin{split}
& \psi_{i,j}=\sum_{n=2}^{K}e_{i,j}^{n}. 
\end{split} 
\label{eqn:cal_psi}
\end{equation} 
The first channel corresponds to background and is thus ignored in the aggregation shown in (\ref{eqn:cal_psi}). The map $\psi$ has continuous values in the range [0,1], where higher values indicate higher likelihood of a voxel corresponding to a SOI, while lower values are indicative of background. 

The segmentor $S$ can be trained using the generated pseudo MRI and the associated expert-segmentations available on the source modality $\{x_c^m, y_c\}$ and optimized using cross-entropy losses. \textcolor{black}{The output segmentation  is computed using an \textit{argmax\/} \rm function}. 

 \textcolor{black}{Given a fixed $G_C^M$ and $S$, the optimal $D_{struct}$ at any point in the optimization of (\ref{eqn:Joint_D_psi}) is given by $\frac{p(x_{m},\psi_{m})}{p(x_{m},\psi_{m})+p(x_{c}^{m},\psi_{c}^{m})}$. The global equilibrium is achieved if and only if the joint distribution of $p(x_{c}^{m},\psi_{c}^{m})$ and $p(x_{m},\psi_{m})$ are matched, i.e., $p(x_{c}^{m},\psi_{c}^{m})=p(x_{m},\psi_{m})$.} 
Thus, unlike methods in~\cite{huo2018synseg,zhang2018,jiang2018tumor,zhao2018supervised,chen2019synergistic} that only constrain geometry, ours constrains both organ geometry and appearance.

\begin{figure}[ht]
\centering
\includegraphics[trim=0 0 0 0,clip,width=0.45\textwidth]{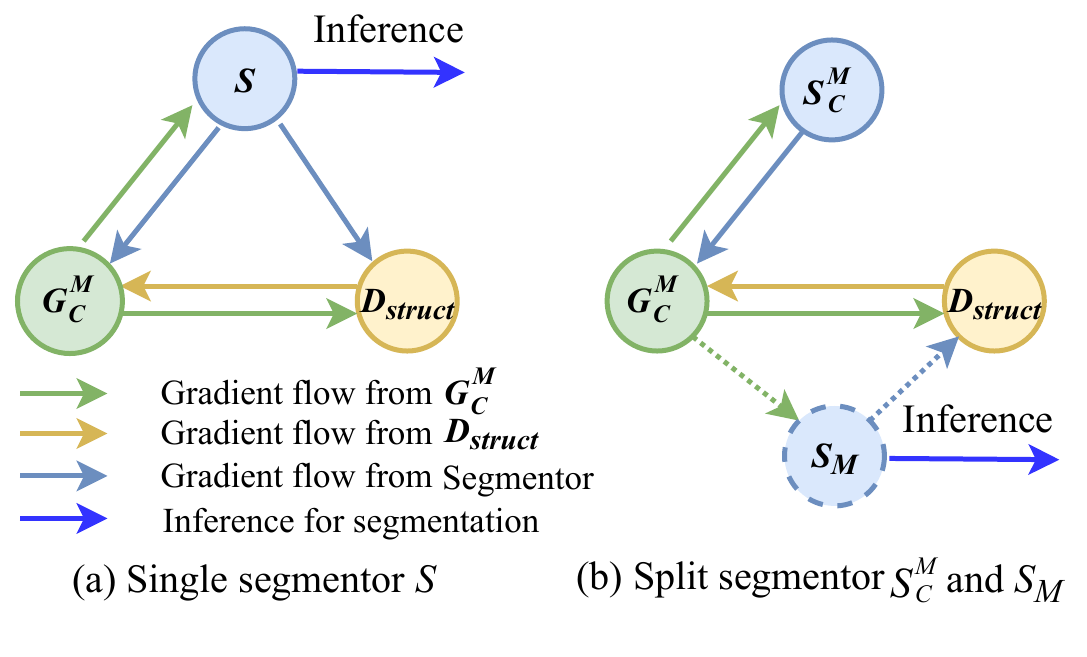}
\caption{\small{\textcolor{black}{Gradient flow between $G_C^M$, $D_{struct}$, and segmentors using (a) single segmentor $S$, and (b) split segmentors $S_C^M$ and $S_M$.} }   \label{fig1:method_gradient_flow}} 
\end{figure}

The joint distribution formulation leads to a cyclical optimization of the generator $G_C^M$ and the  segmentation $S$ networks. Concretely, $S$ requires the output $x_c^m$ of $G_C^M$ for its training, while $G_C^M$ requires the output of $S$, namely $\psi_c^m$ for gradient computation. The gradient flow relation between $D_{struct}$, $G_{C}^M$ and $S$ is illustrated in Fig.~\ref{fig1:method_gradient_flow}(a)\footnote{We only show the components related to our contribution for simpler explanation. Other parts like $D_C$, $G_M^C$, $D_M$ are used as done in CycleGAN.}. 
 
The gradient flow resulting from a network using a single segmentor $S$ is shown in Fig.~\ref{fig1:method_gradient_flow}(a), where the outputs of $G_C^M$ and $S$ are used to update each other. Concretely, $x_c^m$ is used to train $S$, and $\{x_c^m,\psi_c^m\}$ is used to compute the joint distribution  adversarial loss for $G_{C}^M$. As a result, the outputs of $G_{C}^M$ and $S$ can co-adapt to facilitate easy segmentation, but without $x_c^m$ matching the target distribution. In this case, $D_{struct}$ can easily distinguish $p(x_c^m,\psi_c^m)$ from $p(x_m,\psi_m)$, allowing it to reach a stable local minima well before $G_C^M$ reaches its local minima, and possibly prevent GAN convergence. This is because, GAN optimization hinges on achieving Nash equilibrium, whereby all players (generator and discriminator) achieve equal payoffs (reach local minima at similar times).
This would also lower the generalization accuracy of $S$ on real MRI. We address this potential issue by separating the networks used for producing the segmentation probability maps for $G_C^M$ and $D_{struct}$. Thus, segmentation probability maps from network $S_M$ is used to compute $D_{struct}$ loss, and segmentation probability maps  from $S_C^M$ is used for to compute $G_C^M$ loss. The modified configuration of gradient flow is shown in Fig.~\ref{fig1:method_gradient_flow} (b). \\

%\bumpup
\subsection{Split segmentor network}
The split segmentation network consists of two encoder sub-networks, called $E_{M}$ and $E_{C}^{M}$ with a shared decoder network (DE) (Fig.~\ref{fig1:method_overview}), from which two segmentations are generated via $S_{M} = E_{M} \circ DE$ and $S_{C}^{M} = E_{C}^{M} \circ DE$. \textcolor{black}{Shared decoder is used because the high-level contextual features \textcolor{black}{for segmentation} should be same between \textcolor{black}{pseudo and real MRI}. It also reduces the number of parameters required for training $S_M$ and $S_C^M$.} \textcolor{black}{The networks $S_M, S_C^M$ are trained using pseudo MRI data with label ($x_{c}^{m}, y_{c}$) and optimized with cross-entropy losses $L_{seg}^M$ (first part of summation in (\ref{eqn:segmentation loss_share})) for $S_M$ and $L_{seg}^{\bar{M}}$ (second part of summation in (\ref{eqn:segmentation loss_share})) for $S_C^M$, respectively. The overall loss $L_{seg}=L_{seg}^M +L_{seg}^{\bar{M}}$ is computed} as:

\begin{equation}
    \begin{split}
    \setlength{\abovedisplayskip}{1pt}
    \setlength{\belowdisplayskip}{1pt}
    L_{seg} =  &\underbrace{\E_{x_c^m \sim p(x_c^m), y_{c} \sim Y_{c}}[logP(y_{c}|S_M(x_c^m)]}_\text{$L_{seg}^M$}   + 
    \\ &\underbrace{\E_{x_c^m \sim p(x_{c}^{m}), y_{c} \sim Y_{c}}[logP(y_{c}|\textcolor{black}{S_C^M(x_c^m)})]}_\text{$L_{seg}^{\bar{M}}$}. 
    \label{eqn:segmentation loss_share}
    \end{split}
\end{equation}

\textcolor{black}{The network $S_M$ is used to produce $\psi_m$ and $\psi_c^m$ from $x_m$ (solid blue arrow Fig.~\ref{fig1:method_overview}) and $x_c^m$ (solid orange arrow Fig.~\ref{fig1:method_overview}), respectively to compute the gradients for $D_{struct}$. Gradient flow with respect to $G_C^M$, $S_M$, and $D_{struct}$ is shown in Fig.~ \ref{fig1:method_gradient_flow} (b). The modified loss $L_{struct}^D$ \textcolor{black}{to optimize $D_{struct}$}, obtained by replacing $S$ in (\ref{eqn:Joint_D_psi}) with $S_M$ is computed as}: 
\begin{equation}
\label{eqn:local_D_G_D_share}
   \begin{split}
       &\textcolor{black}{\max\limits_{D_{struct}}} L^{D}_{struct} = \\ & \textcolor{black}{\E_{(x_m,\psi_m) \sim p(x_m, S_M(x_m))}  [log(D_{struct}([ x_m,\psi_m]))] } + \\ & \textcolor{black}{\E_{(x_c^m,\psi_c^m) \sim p(x_{c}^{m},S_M(x_{c}^{m}))} [log(1-(D_{struct}([x_{c}^{m},\psi_{c}^{m}])))]}. 
\end{split} 
\end{equation}	
\textcolor{black}{The network $S_C^M$ is used to produce $\bar{\psi}_c^m$ from $x_c^m$ (dotted orange arrow Fig.~\ref{fig1:method_overview}) in order to compute the gradients for $G_C^M$. \textcolor{black}{We use $\bar{\psi}_c^m$ to differentiate from  $\psi_c^m$ that is generated using $S_M$ in (\ref{eqn:local_D_G_D_share}).} The loss $L_{struct}^G$ \textcolor{black}{to optimize $G_C^M$} is now computed by replacing $S$ in (\ref{eqn:Joint_G_psi}) with  $S_C^M$ \textcolor{black}{ and  $x_c^m$ = $G_C^M$($x_c$)}}: %fool the structure discriminator ($D_{struct}$) and update  the generator $G_C^M$:
\begin{equation}
\begin{split}
&\textcolor{black}{\min\limits_{G_C^M}} \ L^{G}_{struct}= \\  &\textcolor{black}{\E_{(x_c^m, \bar{\psi}_c^m) \sim p(x_c^m,S_C^M(x_c^m))} [1 - log(D_{struct}([x_c^m,\bar{\psi}_c^m]))].} 
\end{split} 
\label{eqn:local_G_G_D_share}
\end{equation}	
The network $S_M$, \textcolor{black}{which is never used in generator update} is used to segment target datasets at test time. \textcolor{black}{We found that this network produces more accurate MRI segmentations than the $S_C^M$ as shown in results}. 

\subsection{Additional losses for unpaired mapping}
In order to train with unpaired CT and MR datasets, \textcolor{black}{we enforce consistent reverse mapping by computing adversarial penalties from target to source generator $G_M^C$ and global \textcolor{black}{intensity} discriminator $D_C$}. This loss is expressed as:
\begin{equation}
\begin{split}
%\scalemath{0.7}
& \textcolor{black}{\max\limits_{D_C}\min\limits_{G_M^C}} \ \textcolor{black}{L^{M\rightarrow C}_{adv}(G_{M}^{C}, D_{C}) }= \textcolor{black}{\E_{x_c \sim p(x_c)}} [log(D_{C}(x_c))] \\ + &  \textcolor{black}{\E_{x_m \sim p(x_m)}} [log(1-(D_{C}(G_{M}^{C}(x_m))]. 
\end{split} 
\label{eqn:over_adversary_loss_C}
\end{equation}
The global adversarial loss is then computed as $L_{adv} = L_{adv}^{C\rightarrow M}+L_{adv}^{M \rightarrow C}$. 
Cyclical consistency loss~\cite{zhu2017unpaired} is computed to account for lack of any pixel-level correspondence between the source and the target domain images in the I2I translation (\textcolor{black}{$G_{C\circlearrowleft M} = G_{M}^{C}(G_{C}^M(x_c))$; $G_{M\circlearrowleft C} = G_{C}^{M}(G_{M}^{C}(x_m))$}) as:
\begin{equation}
	\begin{split}
	   \textcolor{black}{L_{cyc}(G_{C}^M, G_{M}^C)}   =   &\textcolor{black}{\mathbb{E}_{x_c \sim p(x_c)}}\left[\left\|G_{C\circlearrowleft M}(x_c) - x_c\right\|_{1}\right] + \\ &  
 \textcolor{black}{\mathbb{E}_{x_m \sim p(x_m)}}\left[\left\|G_{M\circlearrowleft C}(x_m) - x_m\right\|_{1}\right]. 
 \end{split}
\label{eqn:Cycle}
\end{equation}
The total loss is expressed as: 
\begin{equation}
\setlength{\abovedisplayskip}{1pt}
\setlength{\belowdisplayskip}{1pt}
L_{total}=L_{adv}+\lambda_{cyc}{L_{cyc}}+\lambda_{struct}{L_{struct}} +\lambda_{seg}{L_{seg}},  
\label{eqn:Total loss}
\end{equation}
where $L_{struct}$ corresponds to either $L^{D}_{struct}$ or $L^{G}_{struct}$ depending on whether this loss is for $D_{struct}$ ((\ref{eqn:local_D_G_D_share})) or $G_C^M$ ((\ref{eqn:local_G_G_D_share})). 	
\begin{algorithm}
    %\scriptsize
	\footnotesize
	%\small
	%\normalsize
	%\DontPrintSemicolon
	\LinesNumbered
	{
	\SetKwInOut{Input}{input}\SetKwInOut{Output}{output}
	 \Input{Source (CT) domain:$\{X_{c},y_{c}\}$, Target (MRI) domain:$\{X_{m}\}$}
	 \Output{Segmentation model $\theta_{S_M}$ for MRI}
	 $\theta_{G}$,$\theta_{D}$,$\theta_{D}$$_{C}^{M}$, $\theta_{S_M}$,$\theta_{S_{C}^{M}}$ $\gets$ initialize \;
	 \For {Iter $\leq$ Maximum Iter}
	  {
	  	
		$X_{c}$, $X_{m}$ $\gets$ sample mini-batch from CT, MRI domains\;
		$L_{adv}$, $L_{cyc}$, $L_{s}$ $\gets$ calculated using (\ref{eqn:over_adversary_loss_M}),(\ref{eqn:over_adversary_loss_C}),(\ref{eqn:Cycle}),(\ref{eqn:segmentation loss_share}) \;	
	
		Compute segmentation probability map $\psi_{c}^{m}$ $\gets$ $S_{C}^{M}$ \;
		
		$L_{struct}^{G}$ $\gets$ calculated using  (\ref{eqn:local_G_G_D_share}) \;
		$\theta_{G}$ $\overset{+}{\gets}$ -$\Delta_{\theta_{G}}$($L_{adv}$+ $\lambda_{cyc}{L_{cyc}}$+ $\lambda_{struct}{L_{struct}^{G}}$+ \textcolor{black}{$\lambda_{seg}$ $L_{seg}^{\bar{M}}$})\;
		
		Compute segmentation probability maps $\psi_{m}, \psi_{c}^{m} \gets S_{M}$\;
		
		$L_{struct}^{D}$ $\gets$ calculated using  (\ref{eqn:local_D_G_D_share}) \;
		$\Delta_{D}$,$\theta_{D_{struct}}$ $\overset{+}{\gets}$ -$\Delta_{\theta_{D}}$($L_{adv}$+$\lambda_{struct}{L_{struct}^{D}}$)\;	
		$L_{seg}$ $\gets$ calculated using  (\ref{eqn:segmentation loss_share})\;		
		$\theta_{S_{M}}, \Delta_{S_{C}^{M}} \overset{+}{\gets} -\Delta_{\theta_{S}}(L_{seg})$\; 	
				
		}
	}
	\caption{\label{algo:algorithm}\small{PSIGAN}}
\end{algorithm}

The generators ($G_{C}^{M}, G_{M}^{C}$), discriminators ($D_{M}$, $D_{C}$, and $D_{struct}$), and segmentors ($S_M, S_C^M$) are updated with the following gradients, $-\Delta_{\theta_{G}}(L_{adv}+ \lambda_{cyc}{L_{cyc}}+\lambda_{struct}L_{struct}^{G}+$\textcolor{black}{$\lambda_{seg}L_{seg}^{\bar{M}})$}, $-\Delta_{\theta_{D,D_{struct}}}$ ($L_{adv}+\lambda_{struct}L_{struct}^{D}$) and $-\Delta_{\theta_{S}}(L_{seg})$, respectively. The algorithm for the proposed method is shown in Algorithm \ref{algo:algorithm}.

\subsection{Implementation and network structure}
All networks were implemented using the Pytorch library and were trained on Nvidia GTX V100 with 16 GB memory. Training was done using ADAM algorithm\cite{kingma2014adam} with an initial learning rate of 1e-4 and batch size of 2. We set $\lambda_{cyc}$=10, $\lambda_{struct}$=0.5 and $\lambda_{seg}$=5 in the training. \textcolor{black}{The appropriate values for these hyper-parameters were selected empirically from the T2w MR parotid dataset set (see Supplementary document Sec. I).} The learning rate was kept constant for the first 30 epochs and decayed to zero in the next 30 epochs.

The generator architectures were adopted from DCGAN~\cite{radford2015unsupervised}. The generators ($G_C^M$ and $G_M^C$) consisted of two stride-2 convolutions, 9 residual blocks and two fractionally strided convolutions with half strides and used \textit{tanh\/} \rm activation following the last convolutional layer to produce output images. The discriminator networks ($D_{M}$, $D_{C}$ and $D_{struct}$) were composed of 5 convolutional layers with a kernel size of 4 $\times$4 pixels that resulted in feature maps of size 64,128,256,512,1 in each layer of these networks. Discriminators were implemented as $70\times70$ pixels patchGAN~\cite{isola2017image}. 

The segmentation network was based on the U-net~\cite{ronneberger2015u} with batch normalization added after each convolution filter. Feature encoders, $E_M$ and $E_C^M$ of the segmentation network were implemented using 2 successive operations of convolution using kernels of size (3$\times$3), batch normalization and \textit{ReLu\/} \rm activation followed by max pooling. Four max-pooling operations were used in the encoder structure for subsampling the image feature maps leading to the feature sizes of 64, 128, 256 and 512. Skip connections from both the encoder networks layers are used to combine with the decoder layer features to prevent segmentation blurring. The decoder network (DE) was implemented using four unpooling operations to upsample the features back to the original image resolution.

\subsection{Evaluation Metrics}
Per-organ segmentation accuracies were measured using Dice similarity coefficient (DSC), computed from  voxel-wise true positives (TP), false positives (FP) and false negatives (FN) ($DSC = \frac{2\times TP}{FP+2\times TP+FN}$), and Hausdorff Distance at 95$^{th}$ percentile (HD95) as suggested in~\cite{menze2015multimodal}. \textcolor{black}{We also computed an overall DSC that is an average DSC over all the structures segmented in a given scan.}

\subsection{Compared methods}
We compared our method against multiple UDA segmentation methods including the CycadaGAN\cite{hoffman2017cycada}, segmentation structure matching (SA)\cite{tsai2018learning}, segmentation entropy matching (ADVENT)\cite{vu2019advent}, SynSeg-Net\cite{huo2018synseg}, and SIFA\cite{chen2019synergistic}. Also, we compared against the CycleGAN\cite{zhu2017unpaired} with a U-net\cite{ronneberger2015u} and the UNIT\cite{liu2017unsupervised} with a U-net. SynSeg\cite{huo2018synseg} and SIFA\cite{chen2019synergistic} are methods that have been developed and applied to medical image segmentation, while all other compared methods were developed for analyzing natural images. Both SA and ADVENT methods employed additional segmentation matching based adversarial losses for domain adapted segmentation training. \textcolor{black}{We also computed the performance on a model trained with only CT (source) images without domain adaptation. Moreover, we compared performance against supervised MRI segmentation model to ascertain the accuracy upper limit. \textcolor{black}{The supervised MRI segmentation training was done by using the MRI validation set as the training set}.}

Default implementation of the various networks as available from the authors' were used. All networks, including ours, were trained and tested using identical datasets with the same hyperparameter settings (learning rates and batch size).  
\section{Experiments and Results}
We evaluated multiple organ and tumor segmentation from four different MRI sequences arising from \textcolor{black}{two} external institution (with two MR sequences) and two private and \textcolor{black}{internal institution} datasets. All networks were trained using 256$\times$256 CT and MRI image patches enclosing the SOIs. All MR scans were standardized to remove patient-dependent signal intensity variations~\cite{nyul1999standardizing} \textcolor{black}{and normalized to a range [-1,1] for meaningful computation of global and joint distribution adversarial losses.}

\textcolor{black}{Ablation experiments were done to assess the utility of the proposed structure discriminator ($L_{struct}^D$), the global \textcolor{black}{intensity} discriminator ($L_{adv}$) and cycle losses ($L_{cyc}$). We also evaluated  alternative network design choices for computing the structure adversarial GAN losses by using (a) only segmentation probability maps, (b) joint distribution representation using multi-channel segmentation probability maps, (c) a single encoder-decoder segmentation network, and (d) SOI-specific discriminators.} 

\subsection{MRI abdomen organs dataset}
\subsubsection{Data}
Twenty MRIs (\textcolor{black}{T1-DUAL in-phase and T2w spectral pre-saturation inversion recovery or SPIR}) acquired on a 1.5T Philips machine from the Combined Healthy Abdominal Organ Segmentation (CHAOS) challenge data \cite{CHAOS2019} were used to generate segmentation of liver, left and right kidneys, and spleen. Ten MRIs were used for training (without labels) and validation and the remaining 10 scans were used for testing.  \textcolor{black}{Both MRI sequences were acquired to perform fat suppression. \textcolor{black}{These data sets were acquired using a 1.5T Philips MRI, with an image resolution of 256 $\times$ 256 pixels, slice thickness that ranged between 5.5mm to 9mm (average of 7.84mm).} Additional details of the MR sequences are in supplemental document Sec.I.A.} 

\textcolor{black}{Thirty contrast-enhanced portal venous phase CT scans with expert segmentations were obtained from a completely different dataset \cite{landman2015miccai} for UDA training.} \textcolor{black}{The CT images had a resolution of 512$\times$512 pixels, an in-plane resolution ranging between 0.54mm $\times$0.54mm to 0.98mm $\times$0.98mm, and slice thickness ranging between 2.5mm to 5.0mm.} \textcolor{black}{CT images were cropped to contain only body region and then resampled to  256$\times$256 to have the same resolution as the MRI images.} \textcolor{black}{Following histogram standardization, T1w and T2w MRI were clipped to the range of [0,1136] and [0,1814] using the 95th percentile of the reference MRI intensity values, respectively.} Separate segmentation networks were trained for the T1w and T2w MRI using 256$\times$256 pixels image patches. These patches were extracted from 8000 T1w, 7872 T2w MRI slices, and 14038 CT slices. 

\textcolor{black}{We also evaluated the feasibility of our method to segment T2w MRIs acquired on different scanners with various  repetition times (TR), echo times (TE), with and without fat suppression from external institution TCIA-LIHC~\cite{erickson2016radiology} dataset. Six patients were downselected from a total of 97 patients; exclusion criteria were only CT scans (N=57), absence of T2w FSE/TSE MRI (N=22), motion artifacts (N=6), and large liver tumors(N=6). Table~\ref{tab:my_label} shows the sequence details from \textcolor{black}{the CHAOS and the six patients from TCIA-LIHC dataset.}} 

\begin{table}[]
    \centering
    \setlength\tabcolsep{1pt}
    \caption{T2w MRI scanning parameters used in the analysis. ETL: Echo train length; TE: Echo time; TR; repetition time; FatSup: Fat suppressed; FSE: Fast spin echo; TSE: Turbo spin echo; RT: Respiratory trigger; Nav: navigator; FatSat: Fat saturation}
    \label{tab:my_label}
	\centering
	%\scriptsize
	%\tiny
	\scriptsize
	%\footnotesize
	%\small
	%\normalsize
	%\large
	%\Large
	%\LARGE
	%\huge
	%\Huge
    \begin{tabular}{|c|c|c|c|c|c|c|c|}
        \hline 
        Dataset & Series & Manufacturer & Magnet & ETL & TE ms & TR ms & Flip angle \\ \hline
         CHAOS & SPIR FatSup & Philips & 1.5T & 24 & 70 & 1930 & 90$^{\circ}$ \\
         DD-A4NJ & FSE RT & Siemens & 1.5T & 21 & 89 & 5233 & 180$^{\circ}$\\
         DD-A1ED & FSE FatSat RT & GE & 3T & 12 & 83.90 & 10000 & 90$^{\circ}$\\
         DD-A4NF & TSE Nav & Siemens & 1.5T & 17 & 76 & 3386 & 150$^{\circ}$\\
         DD-A113 & FSE & GE & 1.5T & 10 & 78.9 & 7500 & 90$^{\circ}$\\
         DD-A114 & FSE & GE & 1.5T & 16 & 86.8 & 6000 & 90$^{\circ}$\\
         K7-AAU7 & FSE RT & Philips & 1.5T & 85 & 80 & 563 & 90$^{\circ}$\\ \hline
    \end{tabular}
    
\end{table}

\subsubsection{Volumetric segmentation accuracies}
\begin{figure*}
\centering
\includegraphics[width=1\textwidth]{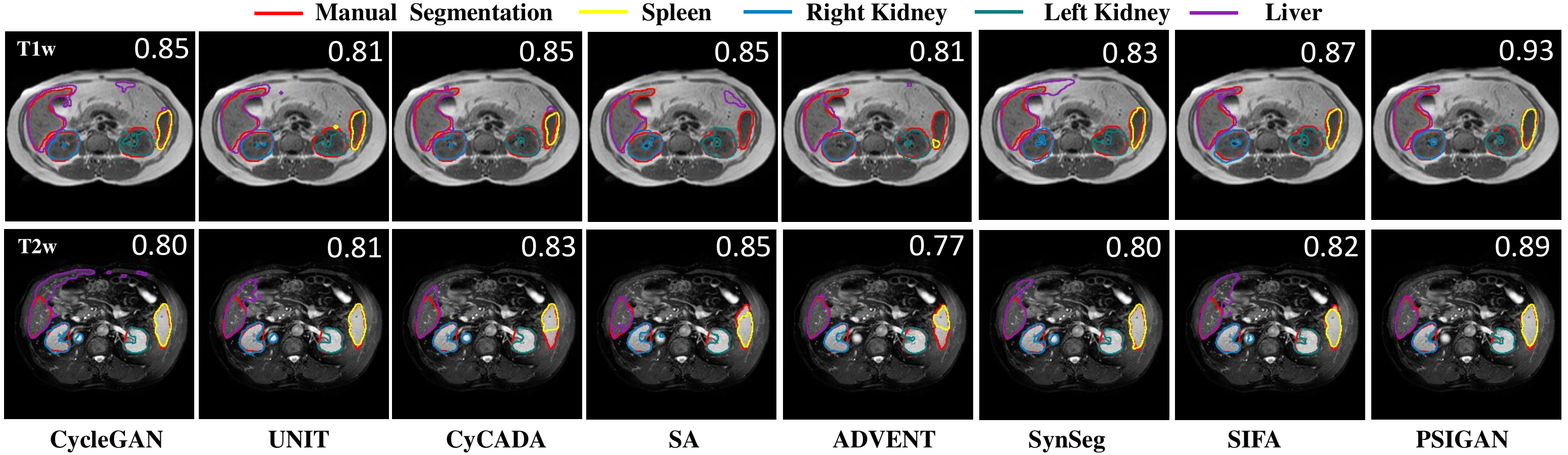}\caption{\small{Segmentation performance of different methods on mdixon T1w and T2w MRI. The overall DSC computed for all organs on this patient is shown in the top right corner of images.}} \label{fig:overlay_T1w_T2w}
\end{figure*}
The volumetric DSC segmentation accuracies for the multiple abdominal organs generated from the T2w and T1w MRI sequences on the testset are shown in Table \ref{tab:adobmen_T1_T2_merge}. \textcolor{black}{ A model trained with only CT images (w/o adaptation) was unable to produce clinically usable segmentations on MRI.} \textcolor{black}{PSIGAN achieved better overall average (computed over all organs) DSC of 0.90 on T2w and 0.87 on T1w MRI and a HD95 of 7.80mm on T2w and 6.90mm on T1w MRI than all other methods except supervised MRI segmentation. PSIGAN accuracy was only slightly lower than the fully supervised MRI segmentation.} 

Fig.~\ref{fig:overlay_T1w_T2w} shows example segmentations generated by the various methods from T1w and T2w MRI. \textcolor{black}{Segmentations produced without adaptation and supervised method are shown in Supplementary Fig.2}. As shown, the PSIGAN segmentations are nearly indistinguishable from the expert's segmentations. 
 \textcolor{black}{\subsubsection{Evaluation on TCIA-LIHC dataset}
\textcolor{black}{PSIGAN produced an average DSC accuracy of 0.87 and an average HD95 accuracy of 8.59mm on this dataset, which is close to that achieved on the CHAOS dataset (Table.~\ref{tab:adobmen_T1_T2_merge})}. PSIGAN segmentations were highly similar to expert delineations on all six cases with highly varying MR tissue contrasts (Supplementary \textcolor{black}{Fig. 5}).} 

\subsubsection{\textcolor{black}{MR to CT UDA segmentation}}
\begin{figure*}[h]
\centering
\includegraphics[width=1\textwidth]{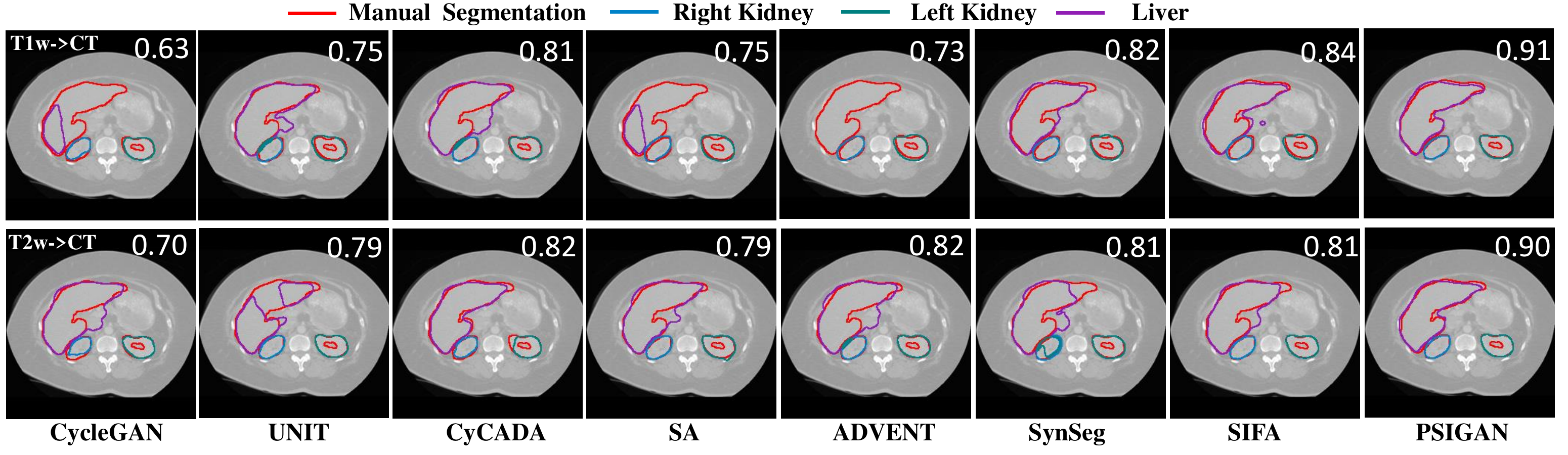}
\caption{\small{Representative segmentations produced by different methods on CT when performing T1w to CT (top row) and T2w to CT UDA segmentation. The overall DSC accuracies for each method are also shown.}} \label{fig:overlay_CT_seg}
\end{figure*}

\textcolor{black}{We also evaluated our and other approaches to segment CT images by using either T1w or T2w MRI as the source modality. Validation was done using 15 CTs and testing was done on remaining 15 cases. Fig.~\ref{fig:overlay_CT_seg} shows segmentations computed on a representative CT case. The corresponding segmentations without adaption and with supervised segmentation are shown in Supplementary Fig. 6. The DSC and HD95 accuracies of all methods computed from the test set are in Supplementary Table I. PSIGAN produced  an overall average DSC of 0.90 and HD95 of 10.35mm when performing T2w to CT UDA and an average DSC of 0.89 and HD95 of 10.30mm when performing T1w to CT UDA segmentation. The next closest method SIFA produced a lower average DSC of 0.86 from T2w to CT and 0.85 from T1w to CT UDA segmentations and higher average HD95 of 14.26mm and 16.20mm for T2w to CT and T1w to CT UDA segmentations, respectively.} 

\begin{table*} 
\centering{\caption{\small{Overall segmentation accuracy on CHAOS dataset. Liver-LV, Spleen-SP, Left kidney-LK, Right kidney-RK. Overall average (Avg) is also shown.}} 
	\label{tab:adobmen_T1_T2_merge} 
	\setlength\tabcolsep{1pt}
	\centering
	%\scriptsize
	%\tiny
	\scriptsize
	%\footnotesize
	%\small
	%\normalsize
	%\large
	%\Large
	%\LARGE
	%\huge
	%\Huge
	\centering
\begin{tabular}{c|c|c|c|c|c|c|c|c|c|c|c|c|c|c|c|c|c|c|c|c|c} 
			\hline
			
			\hline
	\multirow{3}{*}{Method}&\multirow{3}{*}{}&
	 \multicolumn{10}{c}{T2w-SPIR MRI (fat suppressed) (N=10)
	 } & \multicolumn{10}{|c}{T1w-DUAL in phase MRI (fat suppressed) (N=10)}\\ 	 
	\cline{3-22}
	{}&{}  &  \multicolumn{5}{c}{DSC $\uparrow$} & \multicolumn{5}{|c}{HD95 $mm$ $\downarrow$}& \multicolumn{5}{|c}{DSC $\uparrow$} & \multicolumn{5}{|c}{HD95 $mm$ $\downarrow$}\\ 
	\cline{3-22}
	{} &{}& {  \textcolor{white}{A}LV }  & {  \textcolor{white}{A}SP  }& {  \textcolor{white}{A}LK }  & {  \textcolor{white}{A}RK  } & {\textcolor{white}{A}\textcolor{black}{Avg.} }
	& {  \textcolor{white}{A}LV }  & {  \textcolor{white}{A}SP  }& {  \textcolor{white}{A}LK }  & {  \textcolor{white}{A}RK  } & {\textcolor{white}{A}\textcolor{black}{Avg.}  }
	& {  \textcolor{white}{A}LV }  & {  \textcolor{white}{A}SP  }& {  \textcolor{white}{A}LK }  & {  \textcolor{white}{A}RK  } & {\textcolor{white}{A}\textcolor{black}{Avg.}  }
	& {  \textcolor{white}{A}LV }  & {  \textcolor{white}{A}SP  }& {  \textcolor{white}{A}LK }  & {  \textcolor{white}{A}RK  } & {\textcolor{white}{A}\textcolor{black}{Avg.}  }\\ 
\hline
	\hline
	\multirow{2}{*}{\textcolor{black}{W/o Adaptation}}&{Avg.} &{0.08}&{0.23}&{0.26}&{0.01}&\multirow{2}{*}{\textcolor{black}{0.15}}&{64.52}&{71.12}&{47.48}&{72.70}&\multirow{2}{*}{\textcolor{black}{63.96}}
	&{0.00}&{0.00}&{0.00}&{0.00}&\multirow{2}{*}{\textcolor{black}{0}}&{97.03}&{141.26}&{89.47}&{108.45}&\multirow{2}{*}{\textcolor{black}{109.05}}\\
	\cline{2-6}  \cline{8-11} \cline{13-16} \cline{18-21}  
{}&{Std.} &{	\textcolor{gray}{0.12}}&{	\textcolor{gray}{0.15}}&{	\textcolor{gray}{0.26}}&{	\textcolor{gray}{0.02}}&{}&{	\textcolor{gray}{34.21}}&{	\textcolor{gray}{22.23}}&{	\textcolor{gray}{19.29}}&{	\textcolor{gray}{10.85}}&{}

&{ \textcolor{gray}{0.00} }&{\textcolor{gray}{0.00}}&{\textcolor{gray}{0.00}}&{\textcolor{gray}{0.00}}&{}&{	\textcolor{gray}{36.47}}&{	\textcolor{gray}{42.04}}&{	\textcolor{gray}{50.05}}&{	\textcolor{gray}{25.06}}
\\
	\cline{1-22}
\multirow{2}{*}{\textcolor{black}{Supervised}}&{Avg.} &{0.92}&{0.87}&{0.92}&{0.91}&\multirow{2}{*}{\textcolor{black}{0.91}}&{11.13}&{6.26}&{4.78}&{4.26}&\multirow{2}{*}{\textcolor{black}{6.61}}

	&{0.92}&{0.87}&{0.85}&{0.86}&\multirow{2}{*}{\textcolor{black}{0.88}}&{7.24}&{5.44}&{5.64}&{4.67}&\multirow{2}{*}{\textcolor{black}{5.75}}

	\\
		\cline{2-6}  \cline{8-11} \cline{13-16} \cline{18-21}  
{}&{Std.} &{	\textcolor{gray}{0.03}}&{	\textcolor{gray}{0.07}}&{	\textcolor{gray}{0.03}}&{	\textcolor{gray}{0.03}}&{}&{	\textcolor{gray}{8.31}}&{	\textcolor{gray}{2.19}}&{	\textcolor{gray}{5.34}}&{	\textcolor{gray}{1.68}}&{}

&{\textcolor{gray}{0.04}   }&{\textcolor{gray}{0.04} }&{\textcolor{gray}{0.14}  }&{\textcolor{gray}{0.18}} &{} &{\textcolor{gray}{2.48}  }&{\textcolor{gray}{1.18} }&{\textcolor{gray}{3.28}  }&{\textcolor{gray}{1.90}}&{}\\	

	\hline
	\hline
	\multirow{2}{*}{CycleGAN}&{Avg.} &   {0.86}   &   {0.75} &   {0.88}  &   {0.87}&\multirow{2}{*}{\textcolor{black}{0.84}}  &  {24.59}   &   {13.51} &   {14.26}  &   {6.08}&\multirow{2}{*}{\textcolor{black}{14.61}}
	&{0.82}&{0.83}&{0.63}&{0.61}&\multirow{2}{*}{\textcolor{black}{0.72}}&{10.40}&{9.45}&{10.72}&{20.26}&\multirow{2}{*}{\textcolor{black}{12.71}}

	\\
\cline{2-6}  \cline{8-11} \cline{13-16} \cline{18-21} 
{}&{Std.} &{\textcolor{gray}{0.09}   }&{\textcolor{gray}{0.17} }&{\textcolor{gray}{0.03}  }&{\textcolor{gray}{0.05}  }&{}&{\textcolor{gray}{11.50}  }&{\textcolor{gray}{16.09} }&{\textcolor{gray}{10.36}  }&{\textcolor{gray}{1.74}}&{}
&{ \textcolor{gray}{0.07} }&{\textcolor{gray}{0.05}}&{\textcolor{gray}{0.14}}&{\textcolor{gray}{0.24}}&{}&{\textcolor{gray}{2.93}}&{\textcolor{gray}{10.07}}&{\textcolor{gray}{5.01}}&{\textcolor{gray}{9.16}}&{}\\
	\cline{1-22}
	\multirow{2}{*}{UNIT}&{Avg.} &   {0.87}   &   {0.76} &   \textbf{0.91}  &   {0.88}&\multirow{2}{*}{\textcolor{black}{0.86}} &   {15.85}   &   {15.45} &   {7.47}  &   {5.87} &\multirow{2}{*}{\textcolor{black}{11.16}}&
{0.89}&{0.81}&{0.64}&{0.62}&\multirow{2}{*}{\textcolor{black}{0.74}}&{11.39}&{10.64}&{11.45}&{12.59}&\multirow{2}{*}{\textcolor{black}{11.52}}

	\\
\cline{2-6}  \cline{8-11} \cline{13-16} \cline{18-21} 
{}&{	\textcolor{gray}{Std.} }&{	\textcolor{gray}{0.08}}&{	\textcolor{gray}{0.23}}&{	\textcolor{gray}{0.03}}&{	\textcolor{gray}{0.04}}&{}&{	\textcolor{gray}{9.63}}&{	\textcolor{gray}{15.80}}&{	\textcolor{gray}{7.46}}&{	\textcolor{gray}{1.68}}&{}&
{\textcolor{gray}{0.02}}&{	\textcolor{gray}{0.07}}&{	\textcolor{gray}{0.17}}&{	\textcolor{gray}{0.23}}&{}&{	\textcolor{gray}{6.62}}&{	\textcolor{gray}{7.84}}&{	\textcolor{gray}{5.66}}&{	\textcolor{gray}{5.82}}\\
	\cline{1-22}
	\multirow{2}{*}{CycaDA}&{Avg.} &   {0.88} &   \textcolor{black}{0.77} & {0.86}  &   {0.86}  &\multirow{2}{*}{\textcolor{black}{0.84}}    &   {12.34}&   {12.24} &   {6.86}  &   {8.56}&\multirow{2}{*}{\textcolor{black}{10.00}}&
{0.85}&{0.73}&{0.71}&{0.70}&\multirow{2}{*}{\textcolor{black}{0.75}}&{11.46}&{12.75}&{9.13}&{17.13}&\multirow{2}{*}{\textcolor{black}{12.62}}

	\\
\cline{2-6}  \cline{8-11} \cline{13-16} \cline{18-21} 
{}&{\textcolor{gray}{Std.} }&{\textcolor{gray}{0.08}   }&{\textcolor{gray}{0.18} }&{\textcolor{gray}{0.03}  }&{\textcolor{gray}{0.06} }&{}&{\textcolor{gray}{15.82}   }&{\textcolor{gray}{14.70} }&{\textcolor{gray}{1.43}  }&{\textcolor{gray}{1.80}}&{} &
{\textcolor{gray}{0.04}}&{\textcolor{gray}{0.12}}&{\textcolor{gray}{0.21}}&{\textcolor{gray}{0.19}}&{}&{\textcolor{gray}{9.79}}&{\textcolor{gray}{13.77}}&{\textcolor{gray}{3.10}}&{\textcolor{gray}{4.31}}&{}\\
\cline{1-22}
	\multirow{2}{*}{SA}&{Avg.} &   {0.86}   &   {0.80} &   {0.89}  &   {0.89} &\multirow{2}{*}{\textcolor{black}{0.86}}&   {18.18}   &   {12.32} &   {10.09}  &   {9.14}&\multirow{2}{*}{\textcolor{black}{12.43}}&
{0.89}&{0.73}&{0.72}&{0.79}&\multirow{2}{*}{\textcolor{black}{0.78}}&{11.67}&{11.33}&{10.30}&{10.33}&\multirow{2}{*}{\textcolor{black}{10.91}}

	\\
\cline{2-6}  \cline{8-11} \cline{13-16} \cline{18-21} 
{}&{\textcolor{gray}{Std.} }&{\textcolor{gray}{0.12} }&{\textcolor{gray}{0.07} }&{\textcolor{gray}{0.07}  }&{\textcolor{gray}{0.03} }&{}&{\textcolor{gray}{11.47}    }&{\textcolor{gray}{11.84} }&{\textcolor{gray}{5.25}  }&{\textcolor{gray}{4.67} }&{}&
{\textcolor{gray}{0.03}}&{\textcolor{gray}{0.10}}&{\textcolor{gray}{0.10}}&{\textcolor{gray}{0.10}}&{}&{\textcolor{gray}{2.86}}&{\textcolor{gray}{9.67}}&{\textcolor{gray}{2.62}}&{\textcolor{gray}{2.61}}&{}\\
\cline{1-22}
	\multirow{2}{*}{ADVENT}&{Avg.} &{0.89}&{0.79}&{0.81}&{0.80}&\multirow{2}{*}{\textcolor{black}{0.82}}&{12.58}&{13.76}&{11.50}&{11.65}&\multirow{2}{*}{\textcolor{black}{12.37}}
&   {0.87}   &   {0.84} &   {0.76}  &   {0.79} &\multirow{2}{*}{\textcolor{black}{0.82}}&   {14.55}   &   {11.72} &   {10.22}  &   {10.62}&\multirow{2}{*}{\textcolor{black}{11.78}}\\
\cline{2-6}  \cline{8-11} \cline{13-16} \cline{18-21} 
{}&{\textcolor{gray}{Std.} }&{\textcolor{gray}{0.08}   }&{\textcolor{gray}{0.03} }&{\textcolor{gray}{0.03}  }&{\textcolor{gray}{0.12} }&{}&{\textcolor{gray}{4.81}    }&{\textcolor{gray}{12.21} }&{\textcolor{gray}{2.26}  }&{\textcolor{gray}{5.26}}&{}&
{\textcolor{gray}{0.03}}&{\textcolor{gray}{0.08}}&{\textcolor{gray}{0.06}}&{\textcolor{gray}{0.08}}&{}&{\textcolor{gray}{6.08}}&{\textcolor{gray}{10.32}}&{\textcolor{gray}{1.78}}&{\textcolor{gray}{1.44}}&{}\\
	
	\hline
	\hline
	\multirow{2}{*}{SynSeg}&{Avg.} &   {0.88}   &   {0.77} &   {0.89}  &   {0.85}&\multirow{2}{*}{\textcolor{black}{0.85}}&   {21.04}    &   {12.63} &   {10.23}  & {6.28}&\multirow{2}{*}{\textcolor{black}{12.55}} &
{0.89}&{0.85}&{0.73}&{0.70}&\multirow{2}{*}{\textcolor{black}{0.79}}&{8.98}&{8.76}&{9.13}&{11.95}&\multirow{2}{*}{\textcolor{black}{9.71}}

	\\
\cline{2-6}  \cline{8-11} \cline{13-16} \cline{18-21} 
{}&{\textcolor{gray}{Std.} }&{\textcolor{gray}{0.08}   }&{\textcolor{gray}{0.19} }&{\textcolor{gray}{0.03}  }&{\textcolor{gray}{0.10} }&{}&{\textcolor{gray}{11.97}   }&{\textcolor{gray}{13.08} }&{\textcolor{gray}{9.05}  }&{\textcolor{gray}{2.44}}&{}&
{\textcolor{gray}{0.03}}&{\textcolor{gray}{0.05}}&{\textcolor{gray}{0.09}}&{\textcolor{gray}{0.19}}&{}&{\textcolor{gray}{2.82}}&{\textcolor{gray}{5.01}}&{\textcolor{gray}{3.10}}&{\textcolor{gray}{5.51}}&{}\\
	\cline{1-22}
	\multirow{2}{*}{SIFA}&{Avg.} &{0.89}   &   {0.77} &   {0.90}  &   {0.89} &\multirow{2}{*}{\textcolor{black}{0.86}}&   {19.20}   &   {13.56} &   {7.28}  &   {5.78}&\multirow{2}{*}{\textcolor{black}{11.46}} &
{0.90}&{0.85}&{0.77}&{0.78}&\multirow{2}{*}{\textcolor{black}{0.83}}&{9.55}&{7.45}&{8.35}&{12.67}&\multirow{2}{*}{\textcolor{black}{9.51}}

	\\
\cline{2-6}  \cline{8-11} \cline{13-16} \cline{18-21} 
{}&{\textcolor{gray}{Std.} }&{\textcolor{gray}{0.09}  }&{\textcolor{gray}{0.22}  }&{\textcolor{gray}{0.02}  }&{\textcolor{gray}{0.03}  }&{}&{\textcolor{gray}{13.01} }&{\textcolor{gray}{16.53}}   &  {\textcolor{gray}{1.21}  }&{\textcolor{gray}{1.44} } &{}&
{\textcolor{gray}{0.02}}&{\textcolor{gray}{0.05}}&{\textcolor{gray}{0.12}}&{\textcolor{gray}{0.07}}&{}&{\textcolor{gray}{2.92}}&{\textcolor{gray}{4.02}}&{\textcolor{gray}{1.76}}&{\textcolor{gray}{7.55}}&{}\\
            \hline
			\hline	
	\multirow{2}{*}{PSIGAN}&{Avg.} &   {\textbf{0.91}}   &   {\textbf{0.87}} &   {\textbf{0.91}}  &   {\textbf{0.90}} &  \multirow{2}{*}{\textbf{\textcolor{black}{0.90}}}& {\textbf{11.15}} &    {\textbf{8.34}}   &  {\textbf{6.81}}  &   {\textbf{4.88}} &\multirow{2}{*}{\textbf{\textcolor{black}{7.80}}}&
{\textbf{0.92}}&{\textbf{0.87}}&{\textbf{0.83}}&{\textbf{0.84}}&\multirow{2}{*}{\textbf{\textcolor{black}{0.87}}}&{\textbf{7.41}}&{\textbf{5.87}}&{\textbf{7.62}}&{\textbf{6.70}}&\multirow{2}{*}{\textbf{\textcolor{black}{6.90}}}

	\\
\cline{2-6}  \cline{8-11} \cline{13-16} \cline{18-21} 
{}&{\textcolor{gray}{Std.} }&{\textcolor{gray}{0.03}}  &  {\textcolor{gray}{{0.02}}   }&{\textcolor{gray}{0.03}  }&{\textcolor{gray}{0.06}  }&{}&{\textcolor{gray}{3.38}   }&{\textcolor{gray}{6.67}}  &  {\textcolor{gray}{{4.66}}  }&{\textcolor{gray}{1.35} } &{}&
{\textcolor{gray}{0.02}}&{\textcolor{gray}{0.03}}&{\textcolor{gray}{0.10}}&{\textcolor{gray}{0.06}}&{}&{\textcolor{gray}{2.76}}&{\textcolor{gray}{1.95}}&{\textcolor{gray}{1.78}}&{\textcolor{gray}{1.44}}&{}\\
			\hline
			
			\hline
	\end{tabular}} 
\end{table*}

\begin{figure*}
\centering
\includegraphics[width=1\textwidth]{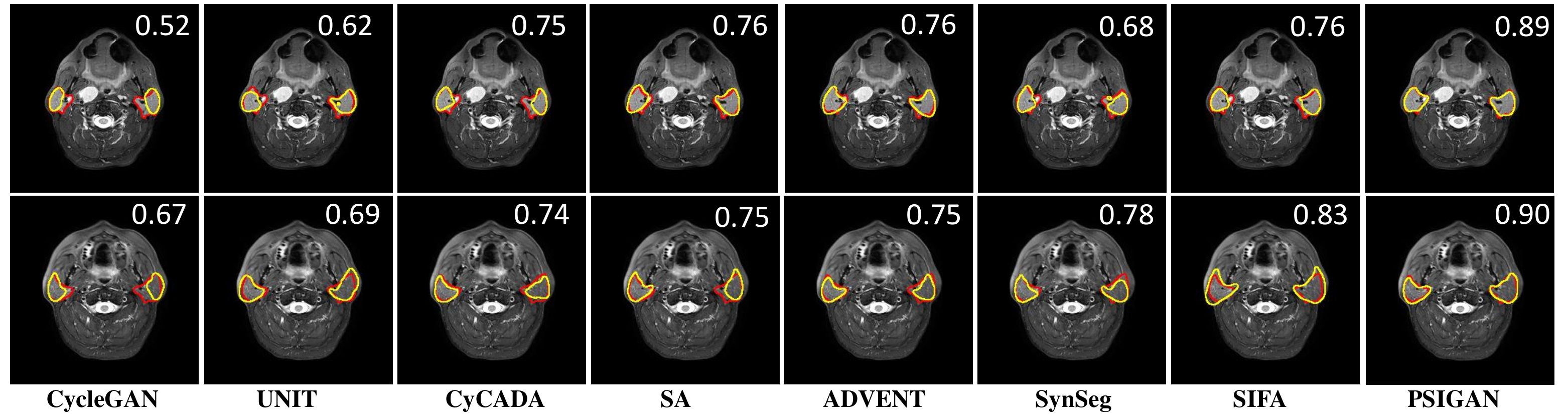}
\caption{\small{Segmentation performance of different methods on T2wFS MRI. Red contour indicates the manual segmentation and the yellow contour indicates the algorithm segmentation. The overall DSC is shown in the top right corner of the images.}} \label{fig:overlay_parotid_2row}
\end{figure*}

\subsection{MRI parotid glands dataset}

\subsubsection{Data}
A private institution head and neck MRI dataset consisting of 162 T2w fat suppressed (T2wFS) head and neck MRIs and obtained from 57 patients who were scanned before and every week during radiation therapy was analyzed. Eighty five MRIs from 30 patients that resulted in 14000 MRI 2D slices were used in training (expert-segmentations removed) and validation (with expert-segmentations added). Remaining 77 MRIs from 27 patients were used for testing. Ninety six head and neck CT scans combining 48 private and 48 open-source public domain database for computational anatomy (PDDCA)~\cite{raudaschl2017evaluation} were used as the source domain. \textcolor{black}{MRI images were clipped to the range of [0,1651] using the 95th percentile of the intensity values of the reference scan following image standardization.}
Two-dimensional patches of size 256 $\times$ 256 pixels containing the head and neck regions \textcolor{black}{obtained after cropping portions outside of the body}, resulting in 15000 CT and 14000 MRI images were used for training. Ablation experiments were done using this dataset, because it had the most MRIs available.

\subsubsection{Volumetric segmentation accuracies}
Table.~\ref{tab:parotid_result} shows the segmentation accuracies for the left and right parotid glands using the various methods \textcolor{black}{including, supervised MRI training and MRI segmentation obtained using a CT network trained without domain adaption.} \textcolor{black}{The CT model trained without any domain adaptation was unable to generate segmentation on MRI.} PSIGAN produced an average DSC of 0.82 and the \textcolor{black}{lowest H95 of 3.06mm, which was slightly lower than supervised method's accuracy with DSC of 0.84 and HD95 of 2.26mm}. The next best  method SIFA reached an average DSC of 0.72 and \textcolor{black}{HD95 of 4.99mm.}  Fig.~\ref{fig:overlay_parotid_2row} shows segmentations generated on a representative test case using the compared methods; \textcolor{black}{the corresponding segmentations produced without adaptation and with the supervised method are in Supplementary Fig. 4.} As seen, PSIGAN segmentations were nearly indistinguishable from the expert contours.

\subsection{MRI lung tumor dataset}
\subsubsection{Data}
A private institution dataset with 75 T2-weighted turbo spin-echo (T2wTSE) MRIs obtained from 27 non-small cell lung cancer (NSCLC) cancer patients scanned before and every week during conventional fractionated external beam radiotherapy of 60 Gy was analyzed. Motion-robust two-dimensional axial images were acquired by using respiratory triggering on a 3T Philips Ingenia scanner (Medical Systems, Best, Netherlands). This is the same dataset as used in our prior work~\cite{jiang2018tumor}. \textcolor{black}{MRI images were clipped to the range of [0,1198] using the 95th percentile of the reference MRI intensity values following image standardization.} Training was done on 9696 unlabeled 2D image patches \textcolor{black}{containing lung tumor} extracted from 35 longitudinal MRI scans of 5 patients, while independent testing was done on the remaining 40 MRI scans from 22 patients. \textcolor{black}{The CT source domain data was obtained from 377 expert-segmented CT scans of NSCLC patients and available from the Cancer Imaging Archive (TCIA)~\cite{aerts2014decoding}. Training was done using 32,000 image patches of size 256$\times$256 pixels \textcolor{black}{containing lung tumor}.} 

\begin{table} 
	\centering{\caption{\small{Overall segmentation accuracy on MRI parotid test dataset. Left parotid - LP, right parotid - RP.}} 
		\label{tab:parotid_result} 
		\setlength\tabcolsep{1pt}
		\centering
		%\scriptsize
		%\tiny
		\scriptsize
		%\footnotesize
		%\small
		%\normalsize
		%\large
		%\Large
		%\LARGE
		%\huge
		%\Huge
		\centering
		\begin{tabular}{c|c|c|c|c|c|c} 
			\hline
			
			\hline			
			\multirow{3}{*}{Method}&
			  \multicolumn{6}{c}{Test T2wFS MRI (N=77)}\\                         			 
			\cline{2-7}
			{} & \multicolumn{3}{c}{DSC $\uparrow$} & \multicolumn{3}{|c}{HD95 $mm$ $\downarrow$}\\ 
			\cline{2-7}
			{} 
			& {   \textcolor{white}{AA}RP \textcolor{white}{AA} }  & {   \textcolor{white}{AA}LP \textcolor{white}{AA}  }& {   \textcolor{black}{Avg.}  } & {   \textcolor{white}{AA}RP \textcolor{white}{AA} }  & {   \textcolor{white}{AA}LP\textcolor{white}{AA}  }& {   \textcolor{black}{Avg.}  } \\ \hline
			\hline
			{\textcolor{black}{W/o Adaption}} &{0.00$\pm$0.00}&{0.00$\pm$0.00}&{\textcolor{black}{0.00}}&{87.81$\pm$18.89}&{85.45$\pm$19.94}&{\textcolor{black}{86.63}}\\
			\hline
			\multirow{1}{*}{\textcolor{black}{Supervised}} &{0.84$\pm$0.06}&{0.84$\pm$0.04}&{\textcolor{black}{0.84}}&{2.24$\pm$0.48}&{2.28$\pm$1.31}&{\textcolor{black}{2.26}}\\
			
			\hline
			\hline
			{CycleGAN} &{0.55$\pm$0.09}&{0.51$\pm$0.11}&{\textcolor{black}{0.53}}&{8.22$\pm$4.81}&{7.38$\pm$2.29}&{\textcolor{black}{3.90}}\\
						\hline
			\multirow{1}{*}{UNIT} &{0.66$\pm$0.06}&{0.62$\pm$0.10}&{\textcolor{black}{0.64}}&{6.91$\pm$5.35}&{5.91$\pm$1.89}&{\textcolor{black}{6.41}}\\
						\hline
			\multirow{1}{*}{CycaDA} &{0.70$\pm$0.08}&{0.69$\pm$0.09}&{\textcolor{black}{0.70}}&{5.59$\pm$2.40}&{4.82$\pm$1.44}&{\textcolor{black}{5.21}}\\
						\hline
				\multirow{1}{*}{SA} &{0.74$\pm$0.07}&{0.71$\pm$0.07}&{\textcolor{black}{0.73}}&{5.05$\pm$1.74}&{4.94$\pm$1.58}&{\textcolor{black}{5.00}}\\
							\hline
						\multirow{1}{*}{ADVENT} &{0.73$\pm$0.12}&{0.71$\pm$0.11}&{\textcolor{black}{0.72}}&{4.66$\pm$2.16}&{5.06$\pm$1.65}&{\textcolor{black}{4.86}}\\
			
			\hline
			\hline
            \multirow{1}{*}{SynSeg} &{0.67$\pm$0.09}&{0.65$\pm$0.09}&{\textcolor{black}{0.66}}&{5.27$\pm$3.56}&{6.08$\pm$2.73}&{\textcolor{black}{5.68}}\\
						\hline
			\multirow{1}{*}{SIFA} &{0.73$\pm$0.08}&{0.71$\pm$0.06}&{\textcolor{black}{0.72}}&{4.95$\pm$1.55}&{5.03$\pm$1.47}&{\textcolor{black}{4.99}}\\

            \hline
		    \hline
            \multirow{1}{*}{PSIGAN} &{\textbf{0.82$\pm$0.03}}&{\textbf{0.81$\pm$0.05}}&{\textbf{\textcolor{black}{0.82}}}&{\textbf{2.98$\pm$1.01}}&{\textbf{3.14$\pm$1.17}}&{\textbf{\textcolor{black}{3.06}}}\\
			\hline
			
			\hline	
		\end{tabular}} 
	\end{table}

\begin{table} 
	\centering{\caption{\small{Segmentation accuracy on T2wTSE MRI (Lung tumor) test set. }} 
		\label{tab:tumor_result} 
		\setlength\tabcolsep{1pt}
		\centering
		%\scriptsize
		%\tiny
		%\scriptsize
		\footnotesize
		%\small
		%\normalsize
		%\large
		%\Large
		%\LARGE
		%\huge
		%\Huge
		\centering
		\begin{tabular}{c|c|c} 
			\hline
			
			\hline	
			\multirow{2}{*}{Method}&
			  \multicolumn{2}{c}{Test (N=40)}\\                         			 
			\cline{2-3}
			{}  &   \multicolumn{1}{c}{\textcolor{white}{AAA}DSC\textcolor{white}{AAA}} & \multicolumn{1}{|c}{HD95 $mm$}\\ 
			
            \hline
			\hline
			{\textcolor{black}{W/o Adaption}} &{0.00$\pm$0.00}&{\textcolor{black}{$\infty$}}\\
			\hline
			{\textcolor{black}{Supervised}} &{0.80$\pm$0.09}&{7.05$\pm$3.66}\\			
			%\cline{3-5}
			%\cline{1-5}
			\hline
			\hline
			{CycleGAN} &{0.64$\pm$0.20}&{17.83$\pm$15.43}\\
			\hline
			{UNIT} &{0.70$\pm$0.16}&{14.31$\pm$12.26}\\
			
			\hline

			{Cycada} &{0.70$\pm$0.18}&{14.17$\pm$12.91}\\	
	        \hline
			{SA}&{0.71$\pm$0.15}&{12.42$\pm$11.38}\\
			\hline			
			{ADVENT} &{0.72$\pm$0.18}&{12.30$\pm$11.93}\\			
			\hline
			\hline		
			{SynSeg} &{0.69$\pm$0.20}&{15.70$\pm$11.98}\\
			\hline
			{SIFA} &{0.72$\pm$0.15}&{11.97$\pm$6.12}\\
			\hline
			{Tumor-aware} &{0.72$\pm$0.16}&{12.88$\pm$11.08}\\

			\hline
			\hline
			{PSIGAN} &{\textbf{0.77$\pm$0.10}}&{\textbf{7.90$\pm$4.40}}\\
			\hline
			
			\hline			
		\end{tabular}} 
	\end{table}

\subsubsection{Volumetric segmentation accuracies}
\begin{figure*}
\centering
\includegraphics[width=1\textwidth]{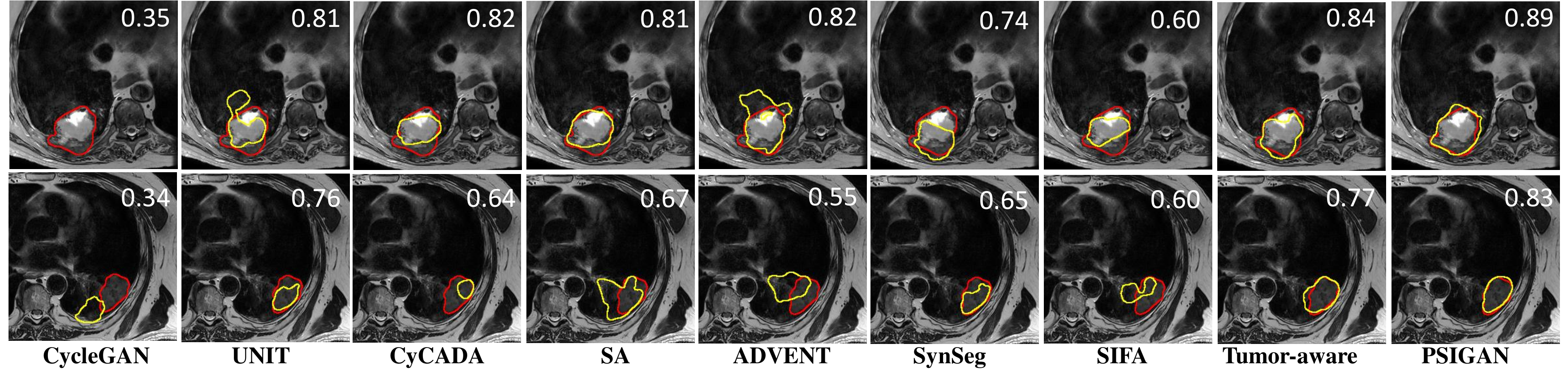}
\caption{\small{MRI lung tumor segmentation with DSC accuracies of multiple methods. Red contour is expert's, yellow is algorithm segmentation.}} \label{fig:tumor_overlay}
\end{figure*}

Table~\ref{tab:tumor_result} shows the lung tumor segmentation accuracies achieved by the various methods.  
The CT segmentation model without domain adaptation failed to generate segmentations on MRI. \textcolor{black}{PSIGAN was more accurate than all except the supervised MRI segmentation method. Both tumor-aware~\cite{jiang2018tumor} and SIFA produced a lower accuracy than PSIGAN.} Fig.~\ref{fig:tumor_overlay} shows two representative examples with algorithms'  segmentation together with expert delineations. \textcolor{black}{The corresponding segmentations without adaption and with supervised segmentation are in Supplementary Fig. 3.}
\subsection{\textcolor{black}{Differences in feature maps extracted using global and joint distribution discriminator}}
\begin{figure*}
\centering
\includegraphics[width=1\textwidth]{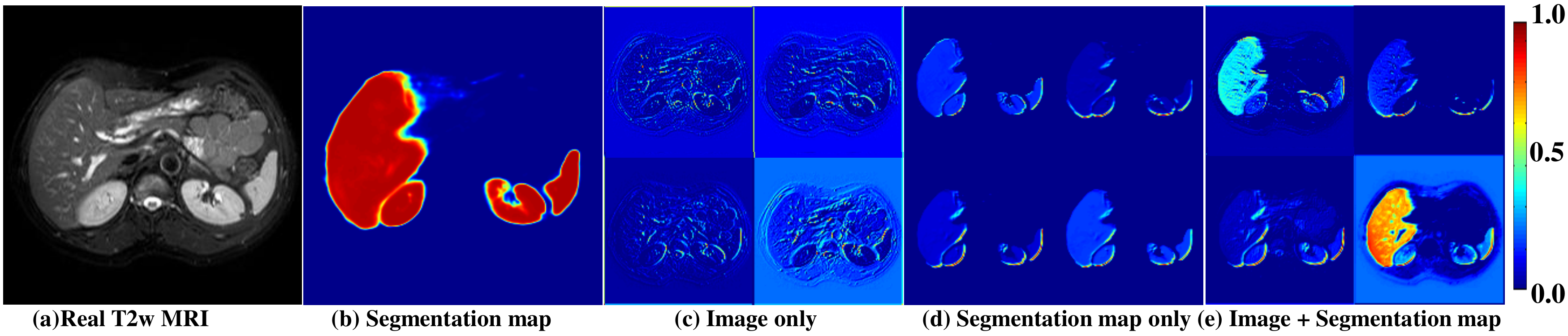}\caption{\small{Feature maps from first convolutional layer and computed with global \textcolor{black}{intensity} discriminator using (c) image, (d) aggregated segmentation probability, and (e) structure discriminator using joint distribution of image and aggregated segmentation probability map. The T2w SPIR MRI (a) and the aggregated segmentation probability map (b) are also shown. \textcolor{black}{Feature response values are normalized to [0,1] for visualization.}}} \label{fig:discriminator_feature_img}
\end{figure*}

\textcolor{black}{Fig.~\ref{fig:discriminator_feature_img} shows four randomly chosen feature maps of the first convolutional layer from the global \textcolor{black}{intensity} discriminator that used only images (Fig.~\ref{fig:discriminator_feature_img}(c)), a discriminator that used only segmentation probability maps (Fig.~\ref{fig:discriminator_feature_img}(d)), and structure discriminator that used a  concatenation of image and aggregated segmentation probability map (Fig.~\ref{fig:discriminator_feature_img}(e)). We visualized the first layer features due to their proximity to the input images and to ascertain what low-level features were relevant for the discriminator. As seen, the image intensity matching global \textcolor{black}{intensity} discriminator extracts features without a clear focus on any part of the image (Fig.~\ref{fig:discriminator_feature_img}(c)), while the segmentation probability matching discriminator amplified features at the SOI boundaries (Fig.~\ref{fig:discriminator_feature_img}(d)), thereby, emphasizing organ geometry. On the other hand, feature responses are higher both within and at SOI boundaries, with a slight emphasis on some background structures (e.g. bottom row  Fig.~\ref{fig:discriminator_feature_img}(e)) when using the structure discriminator. As a result, our method allows the structure discriminator to focus on both geometry and the content within the SOIs, while also preserving some background features.}

\subsection{\textcolor{black}{PSIGAN network design experiments}}
\subsubsection{\textcolor{black}{Structure discriminator}}  \textcolor{black}{We evaluated segmentation performance under the following structure discriminator network designs: (i) multi-channel segmentation probability only, (ii) aggregated segmentation probability map only, (iii)  channel-wise concatenation of image and multi-channel segmentation probability, (iv) SOI specific structure discriminators that used channel-wise concatenation of image and segmentation probability map for each SOI, and (v) default PSIGAN, which used channel-wise concatenation of image and aggregated segmentation probability maps. These tests were done on both the MRI parotid and the T1w SPIR MRI abdomen dataset (CHAOS). Separate networks were trained in each one of these settings from scratch. The main difference between settings (i) and (ii) vs. the rest was the use of segmentation probability only in (i) and (ii) vs. joint-distribution matching of images and segmentation probabilities in (iii), (iv), and (v). Both (iii) and (v) used a single structure discriminator as opposed to $K$ structure discriminators in (iv). Finally, whereas $K$ channels for each organ was used to represent segmentation probability map in (iii), a single aggregated segmentation probability map was used in (v). All other losses including the global \textcolor{black}{intensity} discriminator and cycle consistency losses as used in PSIGAN were used all experiments.} 

\begin{figure*}
\centering
\includegraphics[width=0.9\textwidth]{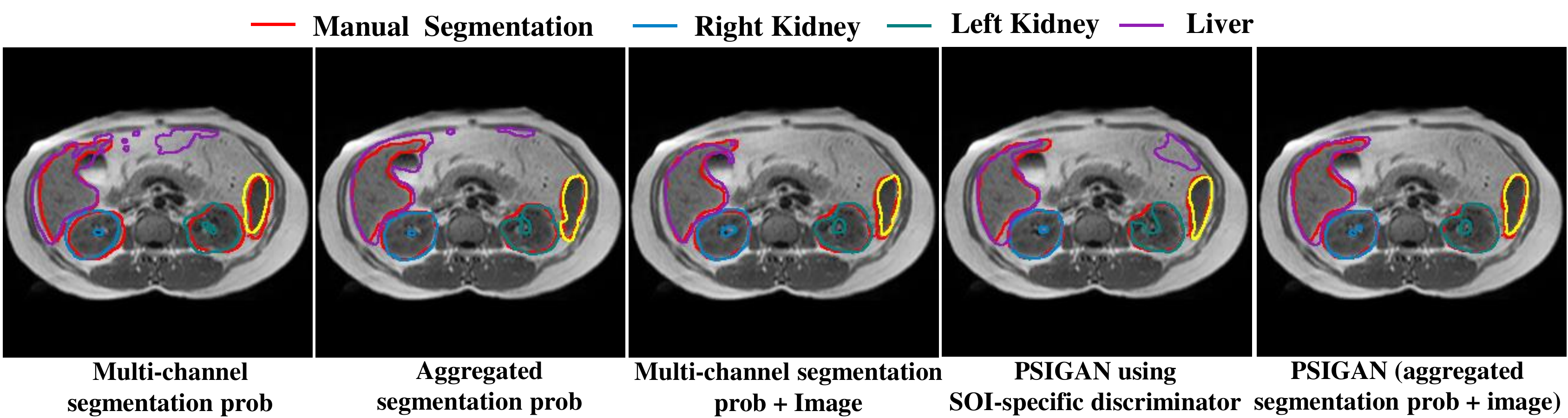}
\caption{\small{Segmentation results for a representative case when using different ways of combining segmentation for computing adversarial loss. SOI-specific discriminator combines segmentation probability map for individual SOI with the images.}} \label{fig:ablation_seg_map_img_abdome}
\end{figure*}

\begin{table*}[ht] 
	\centering{\caption{\small{Impact of using segmentation probability maps for adversarial training on segmentation accuracy.} } 
		\label{tab:ablation_connection_paro_abdomen} 
		\setlength\tabcolsep{1pt}
		\centering
		%\scriptsize
		%\tiny
		%\scriptsize
		\footnotesize
		%\small
		%\normalsize
		
		\centering
		\begin{tabular}{c|c|c|c|c|c|c|c|c} 
		   % \hline
		    
			%\hline 
			
			\hline
			
			\hline
			
			\multirow{3}{*}{Discriminator}&
			 \multicolumn{3}{c|}{T1wFS MR Parotid (N = 77)} & \multicolumn{5}{c}{T1w SPIR MR Abdomen (N = 10)}\\ \cline{2-9}
		    &{\textcolor{white}{A}Parotid right\textcolor{white}{A}}  & {\textcolor{white}{A}Parotid left\textcolor{white}{A}}&{\textcolor{white}{A}\textcolor{black}{Avg.} \textcolor{white}{A}}   & {\textcolor{white}{AA}Liver\textcolor{white}{AA}} & {\textcolor{white}{AA}Spleen\textcolor{white}{AA}} & {\textcolor{white}{A}Kidney left\textcolor{white}{A}} & {\textcolor{white}{A}Kidney right\textcolor{white}{A}}&{\textcolor{white}{A}\textcolor{black}{Avg.}\textcolor{white}{A}}  \\ 
			\hline
			\hline
		    \multirow{1}{*}{i.  Multi-channel segmentation prob} &{0.74$\pm$0.07}&{0.73$\pm$0.06}&{\textcolor{black}{0.74}} & {0.89$\pm$0.02} & {0.84$\pm$0.05} & {0.79$\pm$0.12} & {0.80$\pm$0.07}&{\textcolor{black}{0.83}}\\
		    \cline{1-9}
		    \multirow{1}{*}{ii.  Aggregated segmentation prob } &{0.75$\pm$0.05}&{0.75$\pm$0.06}&{\textcolor{black}{0.75}}& {0.89$\pm$0.03} & {0.83$\pm$0.09} & {0.80$\pm$0.11} & {0.81$\pm$0.09}&{\textcolor{black}{0.83}}\\

			\hline
			\hline

	\multirow{1}{*}{iii.  Multi-channel segmentation prob + Image}    &{0.78$\pm$0.05}&{0.77$\pm$0.05}&{\textcolor{black}{0.78}} & {0.90$\pm$0.03}&{0.85$\pm$0.05}&{0.80$\pm$0.10}&{0.82$\pm$0.07}&{\textcolor{black}{0.84}}\\
	\hline
	\multirow{1}{*}{iv. SOI-specific segmentation prob + Image}& {0.79$\pm$0.03} & {0.78$\pm$0.04} &{\textcolor{black}{0.79}}& {0.91$\pm$0.02} & {0.85$\pm$0.08} & {0.81$\pm$0.10} & {0.82$\pm$0.07} &{\textcolor{black}{0.85}}\\
			\hline
  \multirow{1}{*}{v. Aggregated segmentation prob + Image} &{\textbf{0.82$\pm$0.03}}&{\textbf{0.81$\pm$0.05}}&{\textbf{\textcolor{black}{0.82}}}&{\textbf{0.92$\pm$0.02}}&{\textbf{0.87$\pm$0.03}}&{\textbf{0.83$\pm$0.10}}&{\textbf{0.84$\pm$0.06}}&{\textbf{\textcolor{black}{0.87}}}\\
   
			\hline
			
			\hline
		\end{tabular}} 
\end{table*}		
%\parbreak
\textcolor{black}{As shown in Table~\ref{tab:ablation_connection_paro_abdomen}, the settings (iii), (iv), and (v) produced more accurate segmentations than (i) and (ii). Furthermore, the default setting of PSIGAN that aggregates the segmentation probabilities into a single map produced more accurate segmentation than all other methods. SOI-specific discriminator setting (iv) was similarly accurate as the setting using multi-channel segmentation probability maps in (iii). Segmentations produced on a representative case from the CHAOS dataset by all these methods is shown in Fig.~\ref{fig:ablation_seg_map_img_abdome}. Segmentations on a case from the MR parotid dataset is in \textcolor{black}{Supplementary Fig.9}.} 

\subsubsection{\textcolor{black}{Segmentation network}}
%Impact of single vs. split segmentation network on accuracy}}
We evaluated performance when using single or split segmentor networks (Fig. \ref{fig1:method_gradient_flow}). The structure discriminator and generator losses were computed using (\ref{eqn:Joint_D_psi}) and (\ref{eqn:Joint_G_psi}) for single segmentor, as opposed to using (\ref{eqn:local_D_G_D_share}), (\ref{eqn:local_G_G_D_share}) for the split segmentor case. In addition, we also measured the accuracy of segmentor $S_C^M$ for generating MRI segmentations instead of the default $S_M$ network used in PSIGAN. Both split and single segmentor networks were trained from scratch till convergence with identical training, validation, and testing sets for both parotid and T1w SPIR abdomen organs segmentation. 

The single segmentor configuration was less accurate than the split-segmentator (Table.~\ref{tab:adobmen_T1_T2_merge}, and Table.~\ref{tab:parotid_result}) with an average DSC of 0.85 on the T1w SPIR abdomen segmentation, and an average DSC of 0.80 on the parotid segmentation, respectively. The network parameters were  slightly higher for the split-segmentor (42.97M) compared with single segmentor (38.99M). Adversarial losses for $G_{C}^M$ and $D_{struct}$ during training are shown for split and single network configurations in Supplementary Fig. 10. As shown, the gap in losses for $G_{C}^M$ and $D_{struct}$ stabilized faster for the split than the single segmentor configuration. 

The $S_C^M$ network produced a much lower accuracy than both $S_M$ and single-segmentor networks with an average DSC of 0.84 on the T1w SPIR abdomen segmentation, and an average DSC of 0.77 on the parotid segmentation, respectively. Organ-specific accuracies are in Supplementary Table II. Segmentations on representative examples using these three networks are shown in Supplementary Fig.7 and Fig.8 for both datasets.  

\bumpup
\bumpup
\subsection{\textcolor{black}{Ablation experiments}} 

\subsubsection{Contribution of structure discriminator, global \textcolor{black}{intensity} discriminator, and cycle losses on accuracy}
\textcolor{black}{The goal of this experiment was to evaluate the contribution of each loss on segmentation performance. Both structure and global \textcolor{black}{intensity} discriminator compute an adversarial loss and can train the generator independent of each other. Separate networks were trained from scratch until convergence using identical training, validation, and testing sets from the T2wFS MRI parotid datasets with the following loss settings:
\begin{enumerate}
    \item CT to MR global \textcolor{black}{intensity} discriminator loss $L_{adv}^{C \rightarrow M}$ ( (\ref{eqn:over_adversary_loss_M}) ): Only a global adversarial loss was computed using the discriminator $D_M$ for CT to MRI I2I translation.
    \item CT to MR, MR to CT global \textcolor{black}{intensity} discriminators, and cycle losses $L_{adv}^{C \rightarrow M}$ + $L_{adv}^{M \rightarrow C}$+ $L_{cyc}$  ( (\ref{eqn:over_adversary_loss_M}) , (\ref{eqn:over_adversary_loss_C}), (\ref{eqn:Cycle}) ): Cycle consistency,  global adversarial losses computed using $D_M$ and  $D_C$ (for MR to CT translation) were used. 
    \item Structure discriminator loss $L_{struct}$ ( (\ref{eqn:local_D_G_D_share}), (\ref{eqn:local_G_G_D_share}) ): Joint distribution (image and aggregated segmentation probability map) matching adversarial loss was computed from $D_{struct}$. 
    \item Structure and CT to MR  global \textcolor{black}{intensity} discriminator losses $L_{struct}$ + $L_{adv}^{C \rightarrow M}$ ( (\ref{eqn:local_D_G_D_share}), (\ref{eqn:local_G_G_D_share}), (\ref{eqn:over_adversary_loss_M}) ): Losses from setting 1 and 3 were combined.
    \item Structure, MR to CT global \textcolor{black}{intensity} discriminator, and cycle losses $L_{struct}$ + $L_{adv}^{M \rightarrow C}$ + $L_{cyc}$ ( (\ref{eqn:local_D_G_D_share}), (\ref{eqn:local_G_G_D_share}), (\ref{eqn:over_adversary_loss_C}), (\ref{eqn:Cycle}) ): Cycle loss and loss from setting 3 were combined.
    \item Structure, CT to MR, MR to CT global \textcolor{black}{intensity} discriminators, and cycle losses  $L_{struct}$ + $L_{adv}^{C \rightarrow M}$ +$L_{adv}^{M \rightarrow C}$ + $L_{cyc}$: Default PSIGAN.
\end{enumerate}}
As shown in Table~\ref{tab:ablation_parotid}, $L_{struct}$ loss alone leads to a clear performance improvement compared with the combination of global adversarial and cycle losses ($L_{adv}^{C\rightarrow M}$+ $L_{adv}^{M\rightarrow C}$ + $L_{cyc}$). Addition of either the CT to MR global adversarial (setting 4) or cycle losses (setting 5) to the $L_{struct}$ loss resulted in equivalent  performance improvements. PSIGAN, which combines all the losses produced the most accurate segmentation. Segmentations produced on a representative case by the various networks trained in the aforementioned settings are shown in \textcolor{black}{Supplementary Fig. 11}. 

\subsubsection{Impact of structure discriminator on I2I translation}
\begin{figure}
\centering
\includegraphics[width=0.5\textwidth]{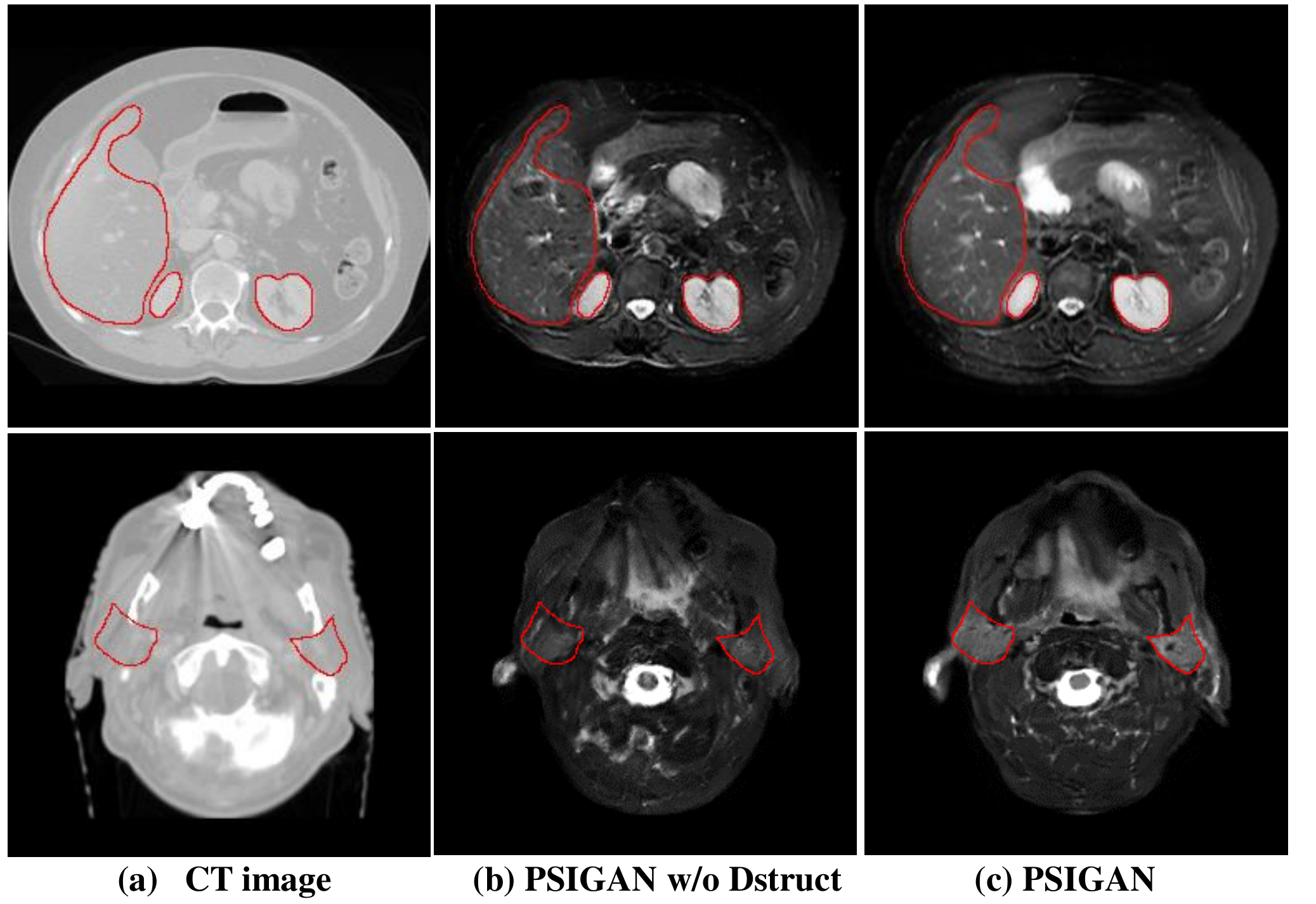}\caption{\small{Impact of $D_{struct}$ on the CT to MRI translation on parotid and abdomen T2w dataset.}} \label{fig:img_translation}
\end{figure}

Fig.~\ref{fig:img_translation} shows I2I translations produced by networks trained without and with structure discriminator on two example images, \textcolor{black}{first one from the T2w SPIR MRI abdomen} and the second from the T2wFS MRI parotid dataset. The source CT image is shown for reference. As shown, the addition of structure discriminator improved the contrast of SOIs with respect to background Fig.~\ref{fig:img_translation}(c) and more accurately modeled the internal characteristics like the regularity in the organization of blood vessels in the liver, as appearing on real MRI. Additional I2I translation results are in \textcolor{black}{Supplementary Fig. 12}. 
	
Quantitative comparison of the distribution of MR signal intensities within the SOIs between the pseudo and real MRIs were computed using Kullback-Leibler (KL) divergence metric\footnote{\textcolor{black}{The metric was computed from pseudo MRI to MRI direction.}}. The method trained without the structure discriminator produced a higher KL-divergence of 1.5 within the parotid glands and 0.14 within liver. Whereas the method trained using structure discriminator produced a K-L divergence of 0.05 for parotid glands and 0.018 for liver.\footnote{We chose liver as this is the largest organ and is highly textured for better quantification of errors in both methods on CHAOS dataset.}

\bumpup
\subsection{Evolution of segmentation probability maps during training}
\begin{figure}
\centering
\includegraphics[width=0.5\textwidth,scale=0.5]{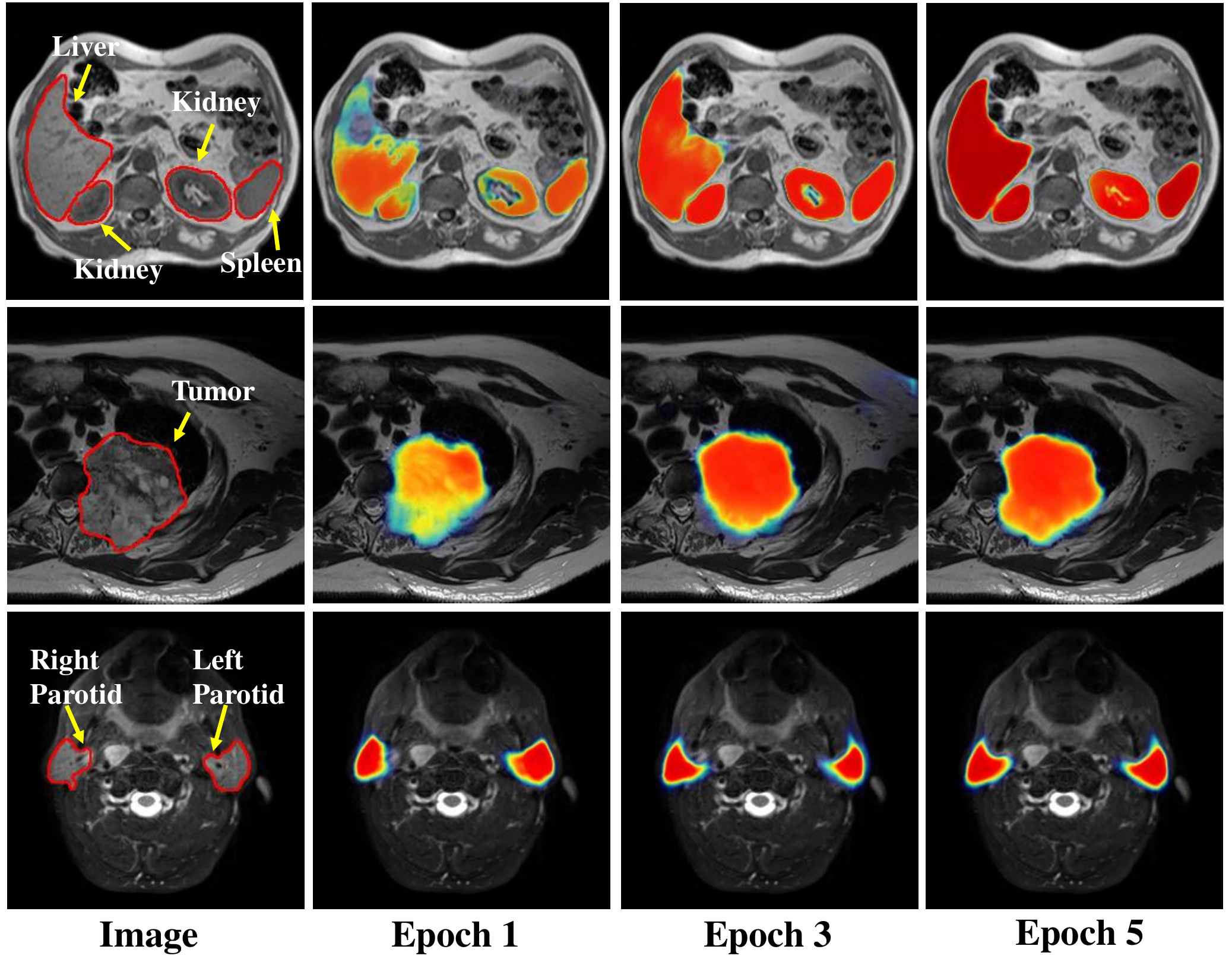}
\caption{\small{Example PSIGAN segmentation probability maps produced during the early training epochs. First column shows expert segmentations on MRI. The structure segmentation probabilities (blue - low probability, and red - high probability) steadily increase with training.}} \label{fig1:attetion_vs_epoch}
\end{figure}

Fig.~\ref{fig1:attetion_vs_epoch} shows the evolution of segmentation probability maps produced on representative examples from the three analyzed datasets during early epochs in training. As seen, the various organs and the tumor are correctly detected and the segmentation probabilities improve, becoming sharper and more focused with training. The higher probabilities are indicated by red color while low probabilities correspond to blue color. These maps clearly indicate their usefulness to constrain I2I translation after training only for a few epochs. 

\begin{table} 
	\centering{\caption{\small{Impact of each component in PSIGAN. RP: Right parotid; LP: Left parotid.}} 
		\label{tab:ablation_parotid} 
		\setlength\tabcolsep{1pt}
		\centering
		%\scriptsize
		%\tiny
		%\scriptsize
		%\footnotesize
		\small
		%\normalsize
		
		\centering
		\begin{tabular}{c|c|c|c|c|c|c} 
		   % \hline
		    
			%\hline 
			
			\hline
			
			\hline
			\multirow{2}{*}{Setting}&\multirow{2}{*}{$L_{adv}^{C \rightarrow M}$}&\multirow{2}{*}{$L_{adv}^{M \rightarrow C}$}&\multirow{2}{*}{$L_{cyc}$}&\multirow{2}{*}{L$_{struct}$}& \multicolumn{2}{c}{DSC}\\ 
			\cline{6-7}
			{}&{}&{}&{}&{}&{RP}&{LP}\\ 
			
			\hline
			\hline	
			
			{1)}&{$\checkmark$}&{$\times$}&{$\times$}&{$\times$} &{0.53$\pm$0.10}&{0.47$\pm$0.08}\\
            \cline{1-7}

			\hline
			
			%\hline

			{2)}&{$\checkmark$}&{$\checkmark$}&{$\checkmark$}&{$\times$}&{0.65$\pm$0.09}&{0.63$\pm$0.10}\\

			\hline
			\hline

			%\hline
			{3)}&{$\times$}&{$\times$}&{$\times$}&{$\checkmark$}&{0.75$\pm$0.06}&{0.74$\pm$0.06}\\			
			%\hline			
			\hline		
	
            {4)}&{$\checkmark$}&{$\times$}&{$\times$}&{$\checkmark$}&{0.77$\pm$0.05}&{0.77$\pm$0.04}\\
            
            \hline
            {5)}&{$\times$}&{$\checkmark$}&{$\checkmark$}&{$\checkmark$}&{0.77$\pm$0.04}&{0.77$\pm$0.06}\\
            
            \hline
            \hline
            
            {6)}&{$\checkmark$}&{$\checkmark$}&{$\checkmark$}&{$\checkmark$}&{\textbf{0.82$\pm$0.03}}&{\textbf{0.81$\pm$0.05}}\\
   
			\hline
			
			\hline		
		\end{tabular}} 
	\end{table}

\section{Discussion}
We introduced PSIGAN, a joint distribution matching method for unsupervised domain adaptation-based multiple organ segmentation. Our method produced highly similar accuracy as supervised method on three different datasets, indicating its ability to learn without requiring target modality labeled image sets. \textcolor{black}{Our method also showed feasibility to segment CT scans when performing T1w to CT and T2w to CT UDA segmentation.}

Joint distribution matching has previously been used to constrain the space of GAN mappings by learning the bi-directional mapping from image to a scalar latent variable~\cite{donahue2016adversarial, li2017ALICE} or from image to vector of class categories~\cite{chongxuan2017triple}. To our knowledge, ours is the first to perform joint distribution matching using pairs of images for UDA segmentation. 

We conducted ablation and network design experiments to determine the utility of the joint distribution matching structure discriminator for both I2I translation and  segmentation. Our results show clear performance improvements when using joint distribution matching adversarial losses. Consistent with prior work that computed adversarial losses using segmentation~\cite{vu2019advent,li2019bidirectional} or with a joint translation-segmentation network~\cite{huo2018synseg,zhang2018}, we also found that features extracted using adversarial discriminators using only the segmentation maps showed a strong preference for SOI geometry, by emphasizing features on SOI boundary. On the other hand, the joint-distribution matching discriminator heavily weighted features corresponding to both the geometry and the appearance within the organs.

We also found that joint distribution formulations that used aggregated segmentation probability maps yielded more accurate segmentations than formulations using either multi-channel segmentation maps or SOI-specific structure discriminators. Performance improvement in the aggregated case could have resulted from increased contextual information from the other organs that was available to the structure discriminator.

Finally, the split segmentor showed a small improvement in accuracy over the single segmentor. However, the choice of the split segmentation network for MRI segmentation clearly impacted accuracy. More specifically, the network that was used for computing discriminator gradients was more accurate than the one used for computing the generator gradients.

Our idea of using segmentation probability maps to guide the translation is similar in principle to attention-guided translation methods~\cite{mejjati2018unsupervised,zhang2019ICML}, which iteratively focus the domain translation network on to regions of interest and produce the desired translation. The main difference is that our method handles simultaneous translation of multiple target and background structures while attention-guided methods are typically restricted to transfiguring a single foreground. Also, as the optimization of the segmentation network is done in a supervised manner, pre-specified image to target relationships can be easily extracted through a fast converging network to constrain translation. Deriving unsupervised attention information on the other hand, would require pre-training of the self-attention network for several epochs before it can be combined with the generator as shown in~\cite{zhang2019ICML}. 

A deficiency of our method as is common to most UDA methods is the inability to handle expert delineation variabilities across modalities that may arise as visibility of structures can vary across modalities. \textcolor{black}{Preliminary evaluation of our method on an external dataset with a variety of MR contrasts and scanning parameters indicates that it is possibly robust to MR contrast variations. However, extensive validation and potential extension to handle large MR contrast variations on much bigger cohorts are needed and is work for future.} Nevertheless, our method outperformed multiple state-of-the-art methods. 

\section{Conclusion}
We developed and evaluated a new unpaired domain adaptation segmentation approach using joint distribution matching structure discriminator for multiple organ segmentation on MRI datasets. Our approach outperformed multiple state-of-the art methods and demonstrated the value of structure discriminator in improving I2I translation and segmentation.

\bibliographystyle{IEEEtran}
\bibliography{mybibliography}

% Generated by IEEEtran.bst, version: 1.14 (2015/08/26)
\begin{thebibliography}{10}
\providecommand{\url}[1]{#1}
\csname url@samestyle\endcsname
\providecommand{\newblock}{\relax}
\providecommand{\bibinfo}[2]{#2}
\providecommand{\BIBentrySTDinterwordspacing}{\spaceskip=0pt\relax}
\providecommand{\BIBentryALTinterwordstretchfactor}{4}
\providecommand{\BIBentryALTinterwordspacing}{\spaceskip=\fontdimen2\font plus
\BIBentryALTinterwordstretchfactor\fontdimen3\font minus
  \fontdimen4\font\relax}
\providecommand{\BIBforeignlanguage}[2]{{%
\expandafter\ifx\csname l@#1\endcsname\relax
\typeout{** WARNING: IEEEtran.bst: No hyphenation pattern has been}%
\typeout{** loaded for the language `#1'. Using the pattern for}%
\typeout{** the default language instead.}%
\else
\language=\csname l@#1\endcsname
\fi
#2}}
\providecommand{\BIBdecl}{\relax}
\BIBdecl

\bibitem{kupelian2014}
P.~Kupelian and J.~Sonke, ``Magnetic-resonance guided adaptive radiotherapy: a
  solution to the future,'' \emph{Semin Radiat Oncol}, vol.~24, no.~3, pp.
  227--32, 2014.

\bibitem{bainbridge2017}
H.~Bainbridge, A.~Salem, R.~Tijssen, M.~Dubec, A.~Wetscherek, E.~C. Van
  \emph{et~al.}, ``Magnetic resonance imaging in precision radiation therapy
  for lung cancer,'' \emph{Transl Lung Cancer Research}, vol.~6, no.~6, pp.
  689--707, 2017.

\bibitem{huo2018synseg}
Y.~Huo, Z.~Xu, H.~Moon, S.~Bao, A.~Assad, T.~K. Moyo \emph{et~al.},
  ``Synseg-net: Synthetic segmentation without target modality ground truth,''
  \emph{IEEE {T}rans. on {M}ed. {I}maging}, vol.~34, no.~4, pp. 1016--1025,
  2018.

\bibitem{kamnistas2017}
K.~Kamnitsas, C.~Baumgartner, C.~Ledig, V.~Newcombe, J.~Simpson, A.~Kane
  \emph{et~al.}, ``Unsupervised domain adaptation in brain lesion segmentation
  with adversarial networks,'' in \emph{Information Processing in Medical
  Imaging}, 2017, pp. 597--609.

\bibitem{zhuTMI2019}
Q.~Zhu, B.~Du, and P.~Yan, ``Boundary weighted domain adaptive neural network
  for prostate \textsc{MR} image segmentation,'' \emph{IEEE Trans. Med
  Imaging}, no.~3, pp. 753--763, 2019.

\bibitem{Zhang_2018_CVPR}
Z.~Zhang, L.~Yang, and Y.~Zheng, ``Translating and {S}egmenting {M}ultimodal
  {M}edical {V}olumes {W}ith {C}ycle- and {S}hape-{C}onsistency {G}enerative
  {A}dversarial {N}etwork,'' in \emph{Proc. IEEE Conf. Comput. Vis. Pattern
  Recognit}, 2018, pp. 9242--9251.

\bibitem{ouyangMICCAI2019}
C.~Ouyang, K.~Kamnistas, C.~Biffi, J.~Duan, and D.~Rueckert, ``Data efficient
  unsupervised domain adaptation for cross-modality image segmentation,''
  \emph{Proc. Int. Conf. Med. Image Comput. Comput.-Assist. Intervent}, pp.
  669--677, 2019.

\bibitem{li2019bidirectional}
Y.~Li, L.~Yuan, and N.~Vasconcelos, ``Bidirectional learning for domain
  adaptation of semantic segmentation,'' in \emph{Proc. IEEE Conf. Comput. Vis.
  Pattern Recognit}, 2019, pp. 6936--6945.

\bibitem{duo2019IEEEAccess}
Q.~Duo, C.~Ouyang, C.~Chen, H.~Chen, B.~Glocker, X.~Zhuang \emph{et~al.},
  ``\textsc{P}n\textsc{P}-\textsc{A}da\textsc{N}et: Plug-and-play adversarial
  domain adaptation network at unpaired cross-modality cardiac segmentation,''
  \emph{{IEEE} Access}, vol.~7, pp. 99\,065--99\,076, 2019.

\bibitem{jiang2018tumor}
J.~Jiang, Y.-C. Hu, N.~Tyagi, P.~Zhang, A.~Rimner, G.~S. Mageras \emph{et~al.},
  ``Tumor-aware, adversarial domain adaptation from \textsc{CT} to \textsc{MRI}
  for lung cancer segmentation,'' in \emph{Proc. Int. Conf. Med. Image Comput.
  Comput.-Assist. Intervent}, 2018, pp. 777--785.

\bibitem{zhu2017unpaired}
J.~Y. Zhu, T.~Park, P.~Isola, and A.~Efros, ``Unpaired image-to-image
  translation using cycle-consistent adversarial networks,'' in \emph{Intl.
  Conf. Computer Vision}.\hskip 1em plus 0.5em minus 0.4em\relax ICCV, 2017,
  pp. 2223--2232.

\bibitem{lee2018diverse}
H.~Y. Lee, H.~Y. Tseng, J.~B. Huang, M.~Singh, and M.-H. Yang, ``Diverse
  image-to-image translation via disentangled representations,'' in \emph{Proc.
  Euro. Conf. Comput. Vis}, 2018, pp. 35--51.

\bibitem{yang2019unsupervised}
J.~Yang, N.~C. Dvornek, F.~Zhang, J.~Chapiro, M.~Lin, and J.~S. Duncan,
  ``Unsupervised domain adaptation via disentangled representations:
  Application to cross-modality liver segmentation,'' \emph{Proc. Int. Conf.
  Med. Image Comput. Comput.-Assist. Intervent}, pp. 255--263, 2019.

\bibitem{zhang2018}
Z.~Zhang, L.~Yang, and Y.~Zheng, ``Translating and segmenting multimodal
  medical volumes with cycle- and shape consistency generative adversarial
  network,'' in \emph{Proc. IEEE Conf. Comput. Vis. Pattern Recognit}, 2018,
  pp. 9242--9251.

\bibitem{bousmalis2017unsupervised}
K.~Bousmalis, N.~Silberman, D.~Dohan, D.~Erhan, and D.~Krishnan, ``Unsupervised
  pixel-level domain adaptation with generative adversarial networks,'' in
  \emph{Proc. IEEE Conf. Computer Vision and Pattern Recognition}, vol.~1,
  no.~2, 2017, pp. 3722--3731.

\bibitem{hoffman2017cycada}
J.~Hoffman, E.~Tzeng, T.~Park, J.-Y. Zhu, P.~Isola, K.~Saenko \emph{et~al.},
  ``{C}y{CADA}: Cycle-consistent adversarial domain adaptation,'' in
  \emph{Proc. Int. Conf. on Machine Learning}, vol.~80, 2018, pp. 1989--1998.

\bibitem{chen2019synergistic}
C.~Chen, Q.~Dou, H.~Chen, J.~Qin, and P.-A. Heng, ``Synergistic image and
  feature adaptation: Towards cross-modality domain adaptation for medical
  image segmentation,'' \emph{arXiv preprint arXiv:1901.08211}, 2019.

\bibitem{salimans2016}
T.~Salimans, I.~Goodfellow, W.~Zaremba, V.~Cheung, A.~Radford, and X.~Chen,
  ``Improved techniques for training gans,'' in \emph{Proc. Adv. Neural Inf.
  Process. Syst}, 2016, pp. 2234--2242.

\bibitem{cheng2018}
C.~Chen, Q.~Dou, H.~Chen, and P.-A. Heng, ``Semantic-aware generative
  adversarial nets for unsupervised domain adaptation in \textsc{C}hest
  {X}-{R}ay segmentation,'' in \emph{Machine Learning in Medical Imaging},
  2018, pp. 143--151.

\bibitem{tsai2019domain}
Y.-H. Tsai, K.~Sohn, S.~Schulter, and M.~Chandraker, ``Domain adaptation for
  structured output via discriminative patch representations,'' in \emph{Proc.
  IEEE Conf. Comput. Vis. Pattern Recognit}, 2019, pp. 1456--1465.

\bibitem{vu2019advent}
T.~H. Vu, H.~Jain, M.~Bucher, M.~Cord, and P.~P{\'e}rez, ``\textsc{ADVENT}:
  \textsc{A}dversarial entropy minimization for domain adaptation in semantic
  segmentation,'' in \emph{Proc. IEEE Conf. Comput. Vis. Pattern Recognit},
  2019, pp. 2517--2526.

\bibitem{li2017ALICE}
C.~Li, H.~Liu, C.~Chen, Y.~Pu, L.~Chen, R.~Henao \emph{et~al.},
  ``\textsc{ALICE}: \textsc{T}owards understanding adversarial learning for
  joint distribution matching,'' \emph{Proc. Adv. Neural Inf. Process. Syst},
  pp. 5495--5503, 2017.

\bibitem{donahue2016adversarial}
J.~Donahue, P.~Kr{\"a}henb{\"u}hl, and T.~Darrell, ``Adversarial feature
  learning,'' \emph{arXiv preprint arXiv:1605.09782}, 2016.

\bibitem{chongxuan2017triple}
L.~Chongxuan, T.~Xu, J.~Zhu, and B.~Zhang, ``Triple generative adversarial
  nets,'' in \emph{Proc. Adv. Neural Inf. Process. Syst}, 2017, pp. 4088--4098.

\bibitem{dou2018unsupervised}
Q.~Dou, C.~Ouyang, C.~Chen, H.~Chen, and P.-A. Heng, ``Unsupervised
  cross-modality domain adaptation of convnets for biomedical image
  segmentations with adversarial loss,'' \emph{arXiv preprint
  arXiv:1804.10916}, 2018.

\bibitem{joyce2018deep}
T.~Joyce, A.~Chartsias, and S.~A. Tsaftaris, ``Deep multi-class segmentation
  without ground-truth labels,'' in \emph{Proc. Int. Conf. Medi. Imag. with
  Deep Learning}, 2018.

\bibitem{dong2018unsupervised}
N.~Dong, M.~Kampffmeyer, X.~Liang, Z.~Wang, W.~Dai, and E.~Xing, ``Unsupervised
  domain adaptation for automatic estimation of cardiothoracic ratio,'' in
  \emph{Proc. Int. Conf. Med. Image Comput. Comput.-Assist. Intervent}, 2018,
  pp. 544--552.

\bibitem{tsai2018learning}
Y.-H. Tsai, W.-C. Hung, S.~Schulter, K.~Sohn, M.-H. Yang, and M.~Chandraker,
  ``Learning to adapt structured output space for semantic segmentation,'' in
  \emph{Proc. IEEE Conf. Comput. Vis. Pattern Recognit}, 2018, pp. 7472--7481.

\bibitem{murez2018image}
Z.~Murez, S.~Kolouri, D.~Kriegman, R.~Ramamoorthi, and K.~Kim, ``Image to image
  translation for domain adaptation,'' in \emph{Proc. IEEE Conf. Comput. Vis.
  Pattern Recognit}, 2018, pp. 4500--4509.

\bibitem{liu2019susan}
F.~Liu, ``\textsc{SUSAN}: segment unannotated image structure using adversarial
  network,'' \emph{Magnetic resonance in medicine}, vol.~81, no.~5, pp.
  3330--3345, 2019.

\bibitem{zhao2018supervised}
H.~Zhao, H.~Li, S.~Maurer-Stroh, Y.~Guo, Q.~Deng, and L.~Cheng, ``Supervised
  segmentation of un-annotated retinal fundus images by synthesis,'' \emph{IEEE
  Trans. on Med. Imaging}, vol.~38, no.~1, pp. 46--56, 2018.

\bibitem{cai2018towards}
J.~Cai, Z.~Zhang, L.~Cui, Y.~Zheng, and L.~Yang, ``Towards cross-modal organ
  translation and segmentation: A cycle-and shape-consistent generative
  adversarial network,'' \emph{Med. Im. Ana.}, vol.~52, pp. 174--184, 2018.

\bibitem{cohen2018}
J.~P. Cohen, L.~Margaux, and H.~Sina, ``Distribution matching losses can
  hallucinate features in medical image translation,'' in \emph{Proc. Int.
  Conf. Med. Image Comput. Comput.-Assist. Intervent}, 2018, pp. 529--536.

\bibitem{armanious2018}
K.~Armanious, C.~Yang, M.~Fischer, T.~K{\"{u}}stner, K.~Nikolaou, S.~Gatidis
  \emph{et~al.}, ``Med\textsc{G}an: {M}edical {I}mage {T}ranslation using
  {GAN}s,'' vol. abs/1806.06397, 2018.

\bibitem{kingma2014adam}
D.~P. Kingma and J.~Ba, ``Adam: A method for stochastic optimization,''
  \emph{Proceedings of the 3rd Int. Conf. on Learning Representations}, 2014.

\bibitem{radford2015unsupervised}
A.~Radford, L.~Metz, and S.~Chintala, ``Unsupervised representation learning
  with deep convolutional generative adversarial networks,'' \emph{arXiv
  preprint arXiv:1511.06434}, 2015.

\bibitem{isola2017image}
P.~Isola, J.-Y. Zhu, T.~Zhou, and A.~A. Efros, ``Image-to-image translation
  with conditional adversarial networks,'' \emph{Proc. IEEE Conf. Comput. Vis.
  Pattern Recognit}, pp. 1125--1134, 2017.

\bibitem{ronneberger2015u}
O.~Ronneberger, P.~Fischer, and T.~Brox, ``U-net: Convolutional networks for
  biomedical image segmentation,'' in \emph{Proc. Int. Conf. Med. Image Comput.
  Comput.-Assist. Intervent}, 2015, pp. 234--241.

\bibitem{menze2015multimodal}
B.~H. Menze, A.~Jakab, S.~Bauer, J.~Kalpathy-Cramer, K.~Farahani, J.~Kirby
  \emph{et~al.}, ``The multimodal brain tumor image segmentation benchmark
  (\textsc{BRATS}),'' \emph{IEEE Trans. on Med. Imaging}, vol.~34, no.~10, p.
  1993, 2015.

\bibitem{liu2017unsupervised}
M.-Y. Liu, T.~Breuel, and J.~Kautz, ``Unsupervised image-to-image translation
  networks,'' in \emph{Proc. Adv. Neural Inf. Process. Syst}, 2017, pp.
  700--708.

\bibitem{nyul1999standardizing}
L.~G. Ny{\'u}l and J.~K. Udupa, ``On standardizing the {MR} image intensity
  scale,'' \emph{Magnetic Resonance in Medicine}, vol.~42, no.~6, pp.
  1072--1081, 1999.

\bibitem{CHAOS2019}
\BIBentryALTinterwordspacing
A.~E. Kavur, M.~A. Selver, O.~Dicle, M.~Barış, and N.~S. Gezer, ``{CHAOS -
  Combined (CT-MR) Healthy Abdominal Organ Segmentation Challenge Data},'' Apr.
  2019. [Online]. Available: \url{https://doi.org/10.5281/zenodo.3362844}
\BIBentrySTDinterwordspacing

\bibitem{landman2015miccai}
B.~Landman, Z.~Xu, J.~Igelsias, M.~Styner, T.~Langerak, and A.~Klein,
  ``\textsc{MICCAI} multi-atlas labeling beyond the cranial vault--workshop and
  challenge,'' 2015.

\bibitem{erickson2016radiology}
B.~Erickson, S.~Kirk, Y.~Lee, O.~Bathe, M.~Kearns, C.~Gerdes \emph{et~al.},
  ``Radiology data from the cancer genome atlas liver hepatocellular carcinoma
  [\textsc{TCGA-LIHC}] collection the,'' \emph{Cancer Imaging Archive}, 2016.

\bibitem{raudaschl2017evaluation}
P.~F. Raudaschl, P.~Zaffino, G.~C. Sharp, M.~F. Spadea, A.~Chen, B.~M. Dawant
  \emph{et~al.}, ``Evaluation of segmentation methods on head and neck ct:
  Auto-segmentation challenge 2015,'' \emph{Medical physics}, vol.~44, no.~5,
  pp. 2020--2036, 2017.

\bibitem{aerts2014decoding}
H.~J. Aerts, E.~R. Velazquez, R.~T. Leijenaar, C.~Parmar, P.~Grossmann,
  S.~Carvalho \emph{et~al.}, ``Decoding tumour phenotype by noninvasive imaging
  using a quantitative radiomics approach,'' \emph{Nature communications},
  vol.~5, p. 4006, 2014.

\bibitem{mejjati2018unsupervised}
Y.~A. Mejjati, C.~Richardt, J.~Tompkin, D.~Cosker, and K.~I. Kim,
  ``Unsupervised attention-guided image-to-image translation,'' in \emph{Proc.
  Adv. Neural Inf. Process. Syst}, 2018, pp. 3693--3703.

\bibitem{zhang2019ICML}
H.~Zhang, I.~Goodfellow, D.~Metaxas, and A.~Odena, ``Self-attention generative
  adversarial networks,'' in \emph{Proc. of Int. Conf. on Machine Learning},
  vol.~97, 2019, pp. 7354--7363.

\end{thebibliography}

\end{document}